\documentclass[aip, reprint]{revtex4-1}
\usepackage{comment}
\usepackage{color}
\usepackage{layout}
\usepackage{ulem}
\usepackage{cancel}
\usepackage{amsmath}
\usepackage{amsfonts}
\usepackage{amssymb}
\usepackage{amsxtra}
\usepackage{amsthm}
\usepackage{mathrsfs}
\usepackage[pdftex]{graphicx}

\draft % marks overfull lines with a black rule on the right

\begin{document}

% Use the \preprint command to place your local institutional report number 
% on the title page in preprint mode.
% Multiple \preprint commands are allowed.
%\preprint{}

\title{$\mathbf{E}\times\mathbf{B}$ drift particle transport in tokamaks} %Title of paper

% repeat the \author .. \affiliation  etc. as needed
% \email, \thanks, \homepage, \altaffiliation all apply to the current author.
% Explanatory text should go in the []'s, 
% actual e-mail address or url should go in the {}'s for \email and \homepage.
% Please use the appropriate macro for the type of information

% \affiliation command applies to all authors since the last \affiliation command. 
% The \affiliation command should follow the other information.

\author{L.A. Osorio-Quiroga}
\author{G.C. Grime}
\email{gabrielgrime@gmail.com}
\affiliation{Institute of Physics, University of São Paulo, São Paulo, 05508-090, Brazil}
\author{M. Roberto}
\affiliation{Physics Department, Aeronautics Institute of Technology, \mbox{São José dos Campos, 12228-900, Brazil}}
\author{R.L. Viana}
\affiliation{Physics Department, Federal University of Paraná, Curitiba, 81531-990, Brazil}
\author{Y. Elskens}
\affiliation{Aix-Marseille Université, UMR 7345 CNRS, PIIM, Marseille, 13397, France}
\author{I.L. Caldas}
\affiliation{Institute of Physics, University of São Paulo, São Paulo, 05508-090, Brazil}
% Collaboration name, if desired (requires use of superscriptaddress option in \documentclass). 
% \noaffiliation is required (may also be used with the \author command).
%\collaboration{}
%\noaffiliation

\begin{abstract}
\textcolor{black}{In tokamaks, modification of the plasma profiles can reduce plasma transport, improving particle confinement. However, this improvement is still not completely understood. In this work, we consider a drift wave test particle model to investigate the influence of the  electric and magnetic field profiles on plasma transport. Test particle orbits subject\textcolor{black}{ed} to \mbox{$\mathbf{E}\times\mathbf{B}$} drift are numerically integrated and their transport coefficient is obtained. We conclude that sheared profiles reduce particle transport, even for high amplitude perturbations. In particular, nonmonotonic electric and magnetic fields produce shearless transport barriers, which are particularly resistant to perturbations and reduce even more the transport coefficient.}
\end{abstract}


% insert suggested PACS numbers in braces on next line

\maketitle %\maketitle must follow title, authors, abstract and \pacs

\section{Introduction}

Tokamaks are one of the most promising devices for achieving commercial thermonuclear fusion power \cite{cowley2016}. \textcolor{black}{However, particle transport limits the plasma confinement. For instance, electrostatic instabilities at the plasma edge produce turbulences leading \textcolor{black}{to} an $\mathbf{E}\times\mathbf{B}$ drift in the particle motion, producing the so-called anomalous transport \cite{horton2018}.}

For some enhanced scenarios of tokamak discharges, the plasma has regions with reduced transport coefficients, known as plasma transport barriers \cite{wolf2002}. Some advances in the understanding of the relation between plasma profiles and the presence of transport barriers were achieved in \cite{cavedon2016}. \textcolor{black}{A transport reduction has been considered due to the $\mathbf{E}\times\mathbf{B}$ shear caused by the electric field profile \cite{burrell1997,connor2004,garbet2004}}. However, these models treat turbulent transport and fluctuation amplitude as a single fluctuating quantity \cite{moyer1995}.

Another approach to describe the onset and destruction of plasma transport barriers regards not only the shear of plasma profiles but also its curvature \cite{kobayashi2017}. Recently, experimental evidence pointed out that, for some regimes, the curvature effects suppress the transport, and this suppression is independent of fluctuations amplitude \cite{kamiya2016}. In this context, models in which the suppression transport is decorrelated \textcolor{black}{from} the amplitude of the fluctuations are needed. One of them is based on shearless transport barriers \cite{caldas2012}.

The shearless transport barriers appear in tokamak plasmas with nonmonotonic electric or magnetic field radial profiles \cite{horton1985,oda1995}. In fact, the location of these transport barriers is associated with the radial position where the plasma \textcolor{black}{profile} shear is null, and it is related to nontwist phenomena in Hamiltonian systems \cite{Carvalho1992,corso1998,diego2000,morrison2000}.

\textcolor{black}{A plasma model to investigate the transport reduction} at the plasma edge was proposed by Horton \cite{horton1998}. This model uses area-preserving maps to describe the motion of test particles subject to the electrostatic drift waves. In this model, particle transport is caused by chaotic orbits that arise in the phase space by the effect of drift wave oscillations.

\textcolor{black}{Some works have considered the influence of the plasma profiles on particle transport regarding Horton's model. Inserting nonmonotonic profiles, nontwist phenomena appear, i.e., a shearless curve acts as a robust barrier preventing the chaotic orbits from escaping the plasma. This nonmonotonicity can be related to the electric \cite{rosalem2014,marcus2019,osorio2021} or magnetic field profiles \cite{mouden2007,grime2023}.}

In this work, we study the influence of \textcolor{black}{such} nonmonotonic profiles on plasma transport using Horton's model. The numerical computations indicate the absence of a shearless transport barrier when monotonic profiles are applied. When a nonmonotonic radial electric field profile is used, the shearless curve appears. In addition, this shearless curve can be destroyed by varying the fluctuation amplitudes, or by modifying the electric and magnetic field profiles. \textcolor{black}{Furthermore, we varied the amplitude of the perturbing modes and obtained the transport coefficient associated with some scenarios of plasma profiles. The results indicate a reduction of the transport in sheared and, especially, in reversed shear scenarios.}

The paper is organized as follows. Section \ref{sec:Drift-wave} is devoted to presenting the particle transport drift wave model used in this work. In section \ref{sec:Influence_profiles} we present the influence of the radial electric and magnetic field profiles in the onset and breakup of shearless transport barriers. In section \ref{sec:Transport_coefficient} we show the transport coefficient analysis. Conclusions are in section \ref{sec:conclusions}.

\section{Drift-wave transport model \label{sec:Drift-wave}}

Plasmas are composed of electrons and ions that move by the effect of electromagnetic fields. A test particle in the plasma moves subject to an electric field $\mathbf{E}$ and a magnetic field $\mathbf{B}$, whose cross product $\mathbf{E}\times \mathbf{B}$ produces an important drift that causes radial transport at the plasma edge in tokamaks \cite{horton2018,ritz1984}.

\begin{figure}[htb]
    \centering
    \includegraphics[width = 0.38\textwidth]{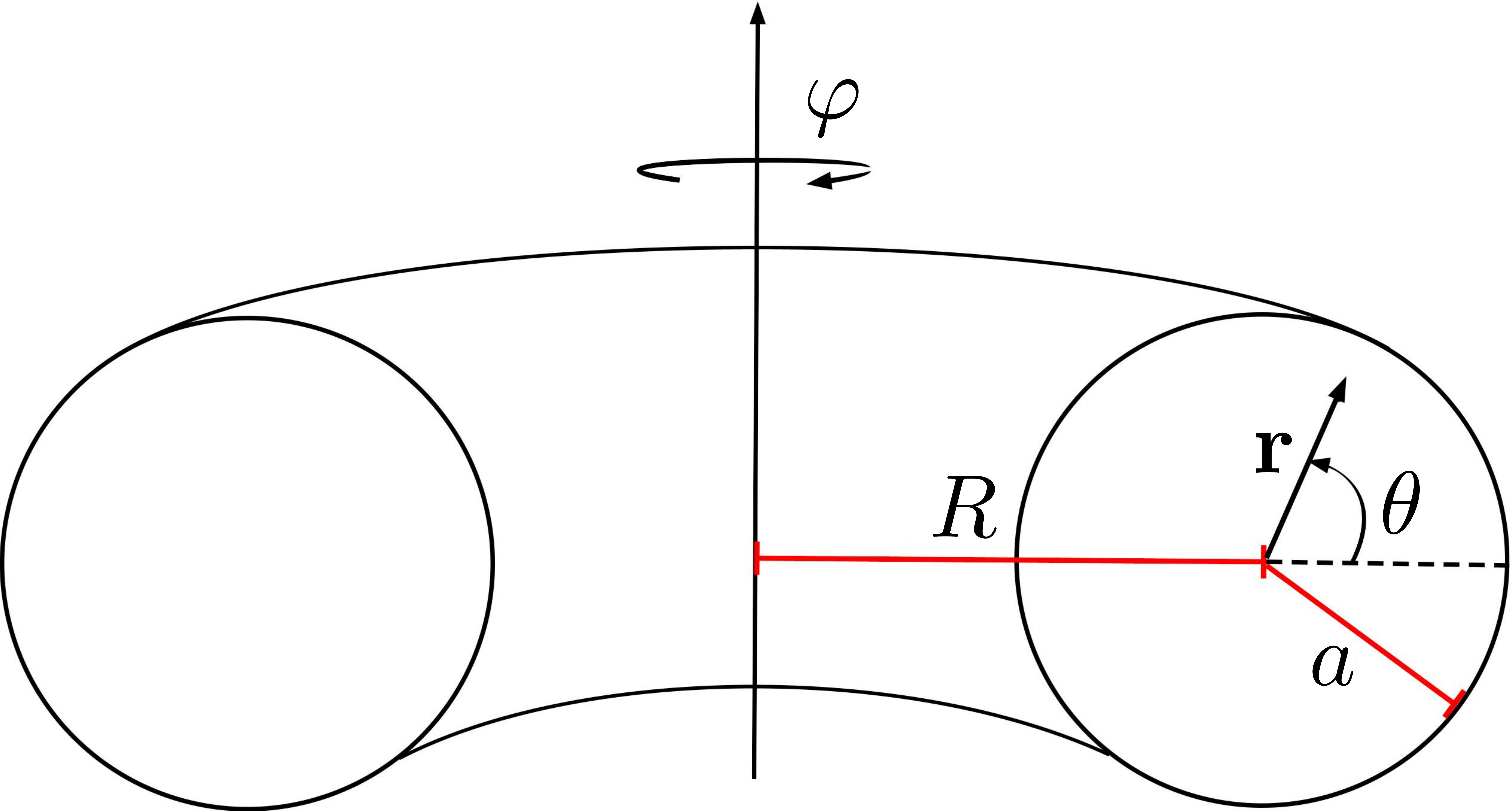}
    \caption{Representation of the tokamak plasma in \mbox{\textcolor{black}{pseudo-toroidal}} coordinates, where $R$ is the major axis and $a$ is the minor axis of the plasma. The \textcolor{black}{pseudo-toroidal} coordinates consist of the toroidal angle $\varphi$, polar angle $\theta$ and radial distance $r$.}
    \label{fig:Toroidal_coordinates}
\end{figure}

Horton's model was proposed to investigate the origins of the transport reduction in reversed shear electric and magnetic field profiles \cite{horton1998}. It is based on the motion of a test particle in the plasma, whose guiding-center moves according to the equation
\begin{equation}\label{eq:horton_model}
    \textcolor{black}{\dot{\mathbf{x}}} = v_{\parallel}\hat{b} + \textcolor{black}{\mathbf{v}_{\mathbf{E}\times\mathbf{B}}, \quad \mathbf{v}_{\mathbf{E}\times\mathbf{B}}=}\dfrac{\mathbf{E}\times\mathbf{B}}{B^2},
\end{equation}
\textcolor{black}{which} has two components: a velocity in the direction of the magnetic field, $\hat{b} = \mathbf{B}/B$, and a drift caused by the cross product $\mathbf{v_{\mathbf{E}\times\mathbf{B}}}$. \textcolor{black}{The dot notation is reserved for total time derivatives, like \mbox{$\dot{\mathbf{x}} = \mathrm{d\mathbf{x}}/\mathrm{d}t$}}. In order to study anomalous transport, other drifts, such as the curvature and gradient of the magnetic field, can be neglected because they have length and time scales larger than the $\mathbf{E}\times\mathbf{B}$ drift. In this limit, the tokamak has a large aspect ratio, i.e., \mbox{$\epsilon = a/R \ll 1$}, where $a$ and $R$ are the minor and major radius of the \textcolor{black}{plasma}, as illustrated in Figure \ref{fig:Toroidal_coordinates}. 

\textcolor{black}{The system of equations \eqref{eq:horton_model} is rewritten into components using} \textcolor{black}{pseudo-toroidal} \textcolor{black}{coordinates}, \textcolor{black}{where} \mbox{$\mathbf{x} = (r,\theta,\varphi)$} \textcolor{black}{is the position vector,} $r$ \textcolor{black}{the radial} component, \textcolor{black}{$\varphi$ and $\theta$} the toroidal and poloidal angles\textcolor{black}{, respectively}, Figure \ref{fig:Toroidal_coordinates}. 

The electric field considered is the sum of an equilibrium and a fluctuating part, such that 
\begin{equation}
    \mathbf{E}(\mathbf{x},t)=\mathbf{E}_r(r)\hat{r} +\tilde{\mathbf{E}}(\mathbf{x},t).
\end{equation}
The fluctuating electric field $\mathbf{\tilde{E}}$ characterizes the electrostatic fluctuations at the plasma edge, originating the anomalous transport. Due to its electrostatic nature, $\nabla\times\mathbf{\tilde{E}}=0$, the fluctuating electric field can be written as the gradient of an electrostatic potential $\tilde{\phi}$, such that, $\mathbf{\tilde{E}}= -\nabla \tilde{\phi}$. Horton's model establishes this potential as a Fourier expansion
\begin{equation}
    \tilde{\phi}(\theta,\varphi,t) = \sum_{n}\phi_n\cos{(M\theta - L\varphi - n\omega_0t - \alpha_n)},
\end{equation}
where $M$ and $L$ are the dominant spatial modes of \textcolor{black}{the fluctuation}, $\omega_0$ \textcolor{black}{its} fundamental \textcolor{black}{angular} frequency, $\phi_n$ and $\alpha_n$ the amplitude and phase of the harmonics, respectively. \textcolor{black}{In general, the electrostatic fluctuations amplitude has a radial dependence\textcolor{black}{;} however, for the purpose of this work, it is considered constant \cite{kwon2000}.}

We consider a plasma equilibrium configuration with the magnetic field given by
\begin{equation}
\mathbf{B}(r) = B_\theta (r)\hat{\theta} + B_\varphi \hat{\varphi}.    
\end{equation}
Since $B_\theta \sim \epsilon B_\varphi$, and \mbox{$\epsilon \ll 1$} for large aspect-ratio tokamaks, we approximate $B \approx B_\varphi \gg B_\theta$. In this limit, the decomposition of Eq. \eqref{eq:horton_model} into components yields
\begin{subequations}\label{eq:motion.components}
\begin{equation}
    \textcolor{black}{\dot{r}} = \dfrac{\textcolor{black}{1}}{Br}\textcolor{black}{\frac{\partial\tilde{\phi}(\theta,\varphi,t)}{\partial\theta}},
\end{equation}
\begin{equation}
    r\textcolor{black}{\dot{\theta}} = \dfrac{rv_{\parallel}\textcolor{black}{(r)}}{Rq(r)} - \dfrac{E_r(r)}{B},
\end{equation}
\begin{equation}
    R\textcolor{black}{\dot{\varphi}} = v_{\parallel}(r),
\end{equation}
\end{subequations}
where
\begin{equation}
    q(r) = \dfrac{rB_\varphi}{RB_\theta(r)}
\end{equation}
is the safety factor profile, another form of writing the radial dependence of the magnetic field.

Introducing the action-like $I=(r/a)^2$ and the angle-like $\psi = M\theta - L\varphi$ variables, and performing the adimentionalization presented in Appendix \ref{sec:appendix}, the equations of motion \eqref{eq:motion.components} \textcolor{black}{reduce} to the equations
\begin{subequations}\label{eq:motion.hamiltonian}
    \begin{align}
        \textcolor{black}{\dot{I}} &= 2M\sum_{n}\phi_n \sin{(\psi - n\omega_0t - \alpha_n)},\\
        \textcolor{black}{\dot{\psi}} &= \epsilon v_{\parallel}\dfrac{M-Lq(I)}{q(I)} - \dfrac{M}{\sqrt{I}}E_r(I),
    \end{align}
\end{subequations}
which form a canonical pair of variables since they satisfy the condition
\begin{equation}
    \dfrac{\partial \dot{\psi}}{\partial \psi} + \dfrac{\partial \dot{I}}{\partial I} = 0,
\end{equation}
which implies the existence of a Hamiltonian function $H$ such that
\begin{equation}
    \dfrac{\mathrm{d}I}{\mathrm{d}t} = -\dfrac{\partial H}{\partial \psi}, \ \ \dfrac{\mathrm{d}\psi}{\mathrm{d}t} = \dfrac{\partial H}{\partial I}.
\end{equation}
Setting
\begin{equation}
\omega(I) = \epsilon v_{\parallel}\dfrac{[M-Lq(I)]}{q(I)} - \dfrac{M}{\sqrt{I}}E_r(I),    
\end{equation}
one can write the Hamiltonian as
\begin{equation}
H(I,\psi,t) = H_0(I) + H_1(\psi,t),    
\end{equation}
where
\begin{subequations}
    \begin{equation}
        H_0(I) = \int^{I}\omega(I') \, \mathrm{d}I'
    \end{equation}
    \begin{equation}
        H_1(\psi,t) = 2M\tilde{\phi}
    \end{equation}
\end{subequations}
are the integrable and perturbative parts of the Hamiltonian system.

If $\phi_n = 0$ for all modes, $I$ \textcolor{black}{will be} a constant of motion and the system \textcolor{black}{will be} integrable \cite{lichtenberg}. \textcolor{black}{Here}, \textcolor{black}{the} perturbative part $H_1$ \textcolor{black}{will be} zero and the \textcolor{black}{dynamics} of the system \textcolor{black}{will be} given by the integrable Hamiltonian $H_0(I)$. In this limit,
\begin{equation}\label{eq:frequency.profile}
    \dot{\psi}/\omega_0 = \omega(I)/\omega_0=\nu(I)
\end{equation}
gives the \textcolor{black}{dimensionless} frequency profile of the orbits. \textcolor{black}{Orbits with} \textcolor{black}{rational} frequency \textcolor{black}{will have a} periodic \textcolor{black}{behaviour}\textcolor{black}{, whereas} orbits \textcolor{black}{with} irrational frequency\textcolor{black}{,} \textcolor{black}{a quasiperiodic one;} the \textcolor{black}{latter} one\textcolor{black}{s} form invariant curves in phase space.

\textcolor{black}{For the perturbed Hamiltonian, $\phi_n\neq 0$,} we have a fluctuating electric field component on the poloidal direction which \textcolor{black}{results in the} \textcolor{black}{radial} drift \mbox{$\mathbf{\tilde{E}_\theta}\times\mathbf{B}_\varphi$}. By the Poincaré-Birkhoff theorem, the rational-frequency orbits form islands \cite{lichtenberg}. Chaotic orbits also appear in phase space and they are bounded by the invariant curves.

\textcolor{black}{After the last invariant curve breaks up, the chaotic trajectories occupy all the accessible phase space and global transport takes place. Moreover, breakup depends on whether the frequency profile has degeneracies, or not.}

According to the KAM theorem, invariant curves whose frequency converges fast to rationals are easier to be destroyed \cite{greene1979}, so, in systems that satisfy the twist condition, i.e.,
\begin{equation}\label{eq:twist.condition}
    \left\lvert\dfrac{\partial \omega(I)}{\partial I}\right\lvert = \left\lvert\dfrac{\partial^2H_0}{\partial I^2}\right\lvert > 0,
\end{equation}
the last invariant curve \textcolor{black}{to be} broken \textcolor{black}{up} has \textcolor{black}{a} frequency value given by the golden mean \cite{meiss1992}. \textcolor{black}{On the other hand}, \textcolor{black}{in} systems that do not satisfy the twist condition \eqref{eq:twist.condition}, called nontwist \textcolor{black}{systems}, invariant curves, which are particularly resistant to periodic perturbations\textcolor{black}{, can appear where KAM theorem does not apply} \cite{viana2021}\textcolor{black}{; these curves are usually called shearless}. \textcolor{black}{Even when} the shearless curve \textcolor{black}{is broken up}, a partial transport barrier still persists \cite{szezech2009}. Hence, the nontwist systems have the so-called shearless transport barriers \cite{morrison2000}.

The model proposed by Horton was introduced to analyze the effects of reversed shear profiles \textcolor{black}{of the electric and magnetic fields} in plasma \cite{horton1998}. In this context, the introduction of nonmonotonic profiles of $E_r(I)$ and $q(I)$ \textcolor{black}{yields} a frequency profile $\omega(I)$ that violates the twist condition \eqref{eq:twist.condition}\textcolor{black}{, resulting in} the presence of shearless transport barriers in phase space.

In the next sections, we will show the effects of applying monotonic and nonmonotonic profiles \textcolor{black}{in} the Hamiltonian system \eqref{eq:motion.hamiltonian}\textcolor{black}. \textcolor{black}{M}ainly, we analyze the onset and breakup of shearless transport curves and the transport in phase space. For  monotonic \textcolor{black}{profiles}, the system satisfies the twist condition. In addition, using a nonmonotonic profile of the electric field, \textcolor{black}{a shearless transport barrier can emerge in} phase space \textcolor{black}{and prevent} the transport of particles at the plasma edge. \textcolor{black}{The} variations of the plasma profiles can cause the \textcolor{black}{breakup} of this shearless curve.

\section{Influence of the plasma profiles \label{sec:Influence_profiles}}

{\color{black}
The Hamiltonian system \eqref{eq:motion.hamiltonian} has one degree of freedom and explicit time dependence with a characteristic angular frequency $\omega_0$. Such time-periodic forced systems are commonly analyzed by a stroboscopic Poincaré section. We numerically integrate the equations \eqref{eq:motion.hamiltonian} and plot the solutions at times $t_j=2\pi j/\omega_0$, for \mbox{$j=0,1,2,\dots , N$}. \textcolor{black}{So}, given an initial condition \mbox{$\mathbf{P}_0=(\psi_0,I_0)$}, we \textcolor{black}{obtain} the next point \textcolor{black}{on the} stroboscopic map, \mbox{$\mathbf{P}_1=(\psi_1,I_1)$}, and all the subsequent ones \textcolor{black}{until having} the orbit \mbox{$\Sigma = (\mathbf{P}_0,\mathbf{P}_1,\mathbf{P}_2,\dots,\mathbf{P}_N)$}. In this work, we applied the Dorman-Prince numerical integrator \cite{prince1981}, with tolerance $10^{-13}$ and maximum \textcolor{black}{stepsize} $10^{-2}$.
}

\begin{figure}[htb]
    \centering
    \includegraphics[width = 0.47\textwidth]{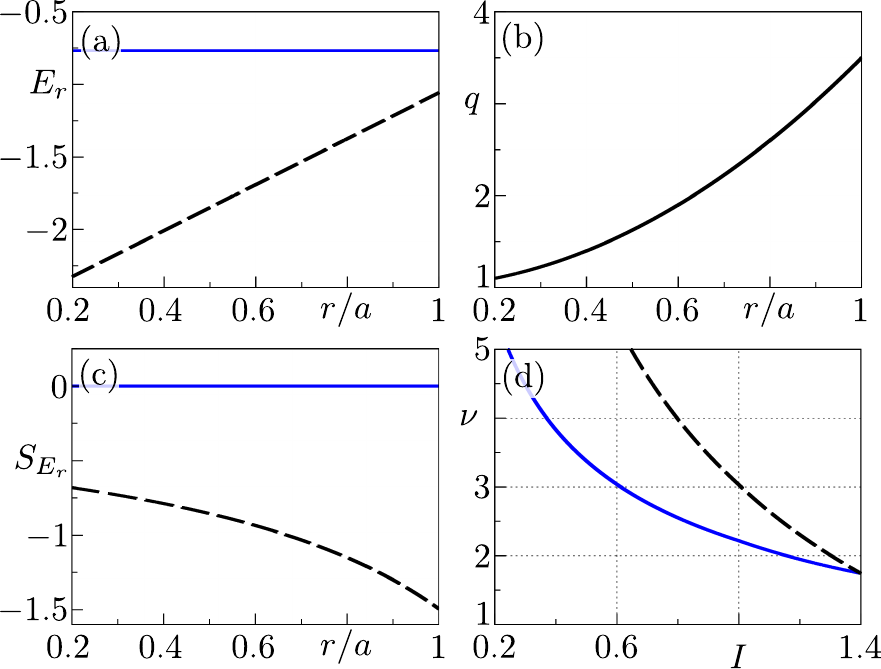}
    \caption{Radial profiles for the (a) equilibrium electric field, $E_r(r)$, (b) safety factor, $q(r)$, (c) electric field shear, $S_{E_r}(r)$ and (d) unperturbed Hamiltonian frequency, $\nu(I)$. The \textcolor{black}{solid} blue line corresponds to the case \mbox{$E_r(r) = \textcolor{black}{\mathrm{constant}}$} and the dashed black line to the \textcolor{black}{linear \mbox{$E_r(r)$} case}.}
    \label{fig:Monotonic_Profiles_Er}
\end{figure}

The results in this section and \textcolor{black}{in section \ref{sec:Transport_coefficient}} are based on parameters and plasma profiles from the TCABR tokamak \cite{nascimento2005}. However, the main conclusions can be generalized to other similar \textcolor{black}{devices}. For TCABR tokamak, the minor and major plasma radii are $a=0.18$ m and \mbox{$R=0.61$ m}, respectively, whereas the minor radius of the vessel chamber is $b = 0.21$ m. The toroidal magnetic field at the plasma center is $B=1.2$ T, the inverse aspect ratio is $\epsilon \approx 0.3$, and the characteristic value of the electric field is regarded as \mbox{$E_0=2.274$ kV/m}. The floating potential parameters are taken based on experimental measurements \textcolor{black}{\cite{marcus2008}}, where for the dominant spatial modes we take $L=4$ and \mbox{$M=16$}, and \mbox{$60$ rad/ms} for the fundamental angular frequency, which corresponds to $\omega_0=5.7$, after carrying out the \textcolor{black}{normalization} presented in Appendix \ref{sec:appendix}. \textcolor{black}{We consider} the phase $\alpha_n = \pi$ for all modes.\par

We are exploring the influence of reversed shear radial profiles of the electric and magnetic fields in plasma transport. Therefore, to avoid introducing the influence of the velocity profile effect, which can affect the twist condition, we assume that the guiding-center of the test particle moves along the magnetic field lines with a constant velocity of \mbox{$3.79$ km/s}, which is consistent with experimental measurements \cite{severo2021} and  corresponds to $v_\parallel = 2.00$, according to the \textcolor{black}{normalization}.

\subsection{Radial electric field}

The considered magnetic field is described by a monotonic safety factor profile given by
\begin{equation}\label{eq:Monotonic-q}
q(r)= \left\{ \begin{array}{ccc}
			q_0+\zeta\left(\dfrac{r}{a}\right)^2 & \mathrm{if} & r \leq a\textcolor{black}{,} \cr\cr
			q_a\left(\dfrac{r}{a}\right)^2 & \mathrm{if} & r > a,
		\end{array}
		\right.
\end{equation}
\noindent where $q_0$ and $q_a$ represent the values at the center and edge of the plasma, and in this case, they are $1.00$ and $3.65$, respectively. The value of $\zeta$ is equal to $q_a-q_0$.

To investigate the influence of the radial equilibrium electric field on plasma transport, we consider: (i) monotonic and (ii) nonmonotonic electric field radial profiles. In both scenarios, we use the same expression for the radial electric field \cite{rosalem2014}, but we adjust the parameters to create different profiles. Specifically, we investigate a constant, a linear, and a parabolic profiles.\par
Then, let us consider the radial electric field profile as the one given by

\begin{equation}\label{eq:electric_field_profile}
E_r(r) = 3\alpha\left(\dfrac{r}{a}\right)^2 + 2\beta\left(\frac{r}{a}\right)+\gamma,
\end{equation}

\noindent where $\alpha$, $\beta$, and $\gamma$ are control parameters that we can adjust.\par

\subsubsection{Monotonic electric field}

\begin{figure}[htb]
    \centering
    \includegraphics[width = 0.45\textwidth]{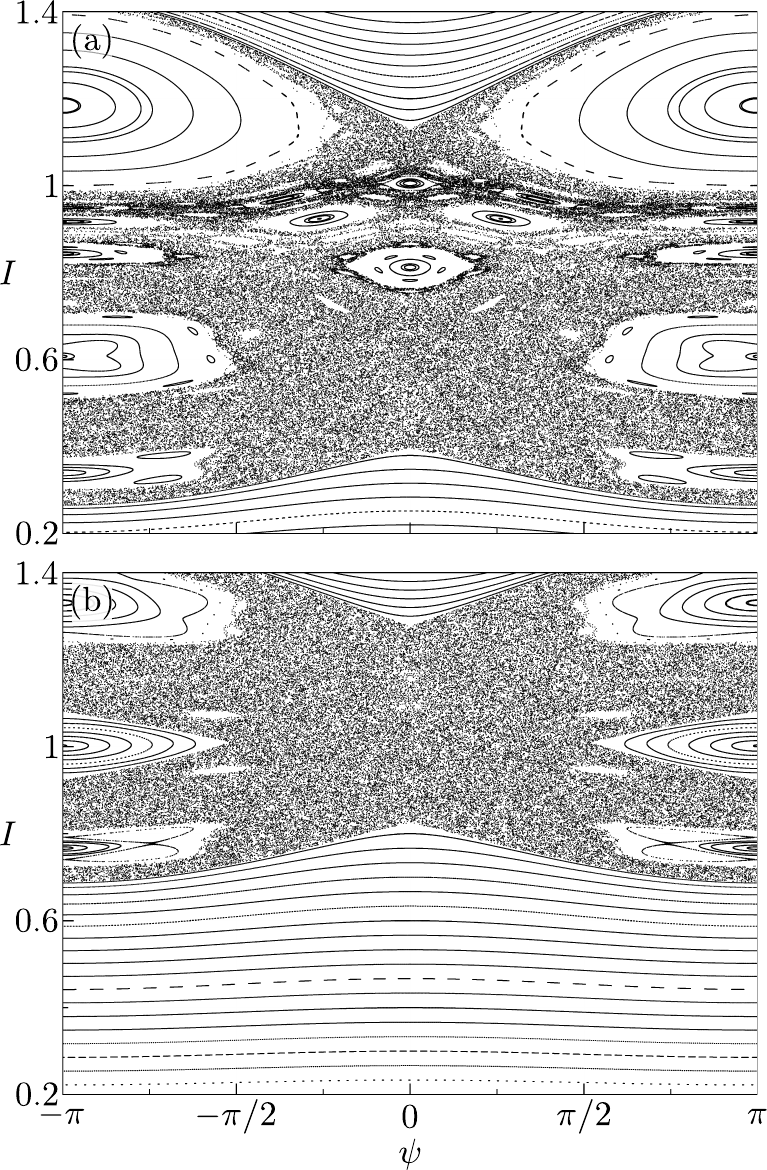}
    \caption{Poincar\'e sections for the (a) constant and (b) linear cases of the equilibrium electric field profile. Chaos suppression occurs when we include a sheared profile for $E_r(r)$.}
    \label{fig:Phase_Space_Er(Monotonic)}
\end{figure}

We consider a constant electric field profile for which \mbox{$\alpha = \beta = 0$ and $\gamma = -0.77$}, and a linear one for which \mbox{$\alpha = 0$}, $\beta = 0.79$ and $\gamma = -2.64$.  We impose both electric fields to be equal at $I=1.4$, a little further than the vessel wall, located at $I_{\mathrm{wall}} = 1.36$.\par

In Figure \ref{fig:Monotonic_Profiles_Er}, we show the monotonic plasma profiles we are considering, the electric field and the safety factor. Furthermore, we show the shear of the electric field
\begin{equation}\label{eq:shear}
    S_{E_r}(r) = \frac{a}{E_r(r)}\frac{\mathrm{d}E_r(r)}{\mathrm{d}r},
\end{equation}
and the frequency profile, $\nu (I)$, obtained from Eq. \eqref{eq:frequency.profile}. \textcolor{black}{We emphasize that the frequency profile depends on the electric and magnetic field profiles. Changes in these profiles modify the frequency of the orbits and, consequently, the phase space configuration.}

For the constant $E_r$, besides a null shear of the electric field, for the unperturbed Hamiltonian, the frequency profile obeys a monotonic behavior, satisfying the twist condition, Figure \ref{fig:Monotonic_Profiles_Er}(d). For this former scenario, we do not expect the appearance of any shearless curve. The same behavior is found for the linear electric field case: although the $S_{Er}$ is not null, this is not enough to make a shearless curve appear, \textcolor{black}{as} $\left\lvert\partial\omega(I)/\partial I\right\lvert>0$ for $I\leq(b/a)^2$. However, the effect of a sheared profile is significant due to it\textcolor{black}{s} changes in the resonance conditions, as we can see from Figure \ref{fig:Monotonic_Profiles_Er}(d). This has a direct influence on chaotic transport, as we will see below.\par

Let us take the resonant modes \mbox{$n=2,3$ and} $4$, and consider for their amplitudes \mbox{$\phi_2=2.95\times 10^{-3}$}, \mbox{$\phi_3=3.66\times 10^{-3}$} and \mbox{$\phi_4=2.08\times 10^{-3}$}. By doing this, we can obtain the Poincar\'e sections for the perturbed Hamiltonian of both cases, the constant and the linear one, and compare which \textcolor{black}{one} is better for the chaotic transport in plasma.\par

In Figure \ref{fig:Phase_Space_Er(Monotonic)}, we present the Poincar\'e sections for the monotonic cases of $E_r(r)$. When $S_{E_r}(r)$ is null across the plasma, it can be seen that the chaotic behavior \textcolor{black}{dominates} in phase space over the regular one. Most of the trajectories can cross the {\color{black}plasma} edge\textcolor{black}{,} at $I = 1.00$, \textcolor{black}{indicating} a\textcolor{black}{n} \textcolor{black}{unsatisfactory} electric field configuration for confinement. On the other hand, a sheared $E_r$ profile suppresses the chaotic behavior and instead establishes \textcolor{black}{more} invariant curves. Due to the three resonant modes we are considering, \textcolor{black}{one observes} in phase space, for the two scenarios, the formation of three main islands of period-one, with centers at $\psi = \pi$. Their radial positions are associated with the resonance condition profile, $\nu(I) = n$. So, when we adjust the electric field from a constant to a linear one, these islands shift their centers and \textcolor{black}{thereby} they change the chaotic dynamics.\par 

Thus, \textcolor{black}{the results indicate} that the \mbox{$E_r$-shear} can reduce the radial chaotic transport in \textcolor{black}{tokamak} plasma\textcolor{black}{s}, which translates into an improvement of the confinement. Nevertheless, we will show below that there are even better scenarios for achieving this purpose. 

\subsubsection{Nonmonotonic electric field}

For now, we indicated that sheared electric field profiles decrease the radial transport in plasmas by suppressing chaotic orbits. However, reversed shear profiles also reduce radial transport and lead to the appearance of robust shearless transport barriers which are more resistant to perturbations than the invariant curves shown in the previous cases \cite{rosalem2014,kwon2000}.\par

In order to investigate the influence of $E_r(r)$ for the nonmonotonic scenarios, let us vary only the parameters $\alpha$ and $\beta$ of \mbox{Eq. \eqref{eq:electric_field_profile}}, keeping the same value of $\gamma$ used for the linear electric field case,  $\gamma = -2.64$. Then, let us keep the same safety factor radial profile, given by \eqref{eq:Monotonic-q}, parallel velocity, and fluctuating potential parameters as the ones already used.\par 

The maximum value of $E_r$ is constrained to \mbox{$E_\mathrm{max} = -0.77$}, which is the same value of $\gamma$ for the constant electric field case. By doing this, we can modify $\alpha$ and $\beta$ at the same time, \textcolor{black}{restraining} the shearless position of the profile to some value $k=r_\mathrm{s}/a = - \beta/(3\alpha)$, where $(\mathrm{d}E_r/\mathrm{d}r)\lvert_{r_{\mathrm{s}}}=0$. \par

\begin{figure}[htb]
    \centering
    \includegraphics[width = 0.47\textwidth]{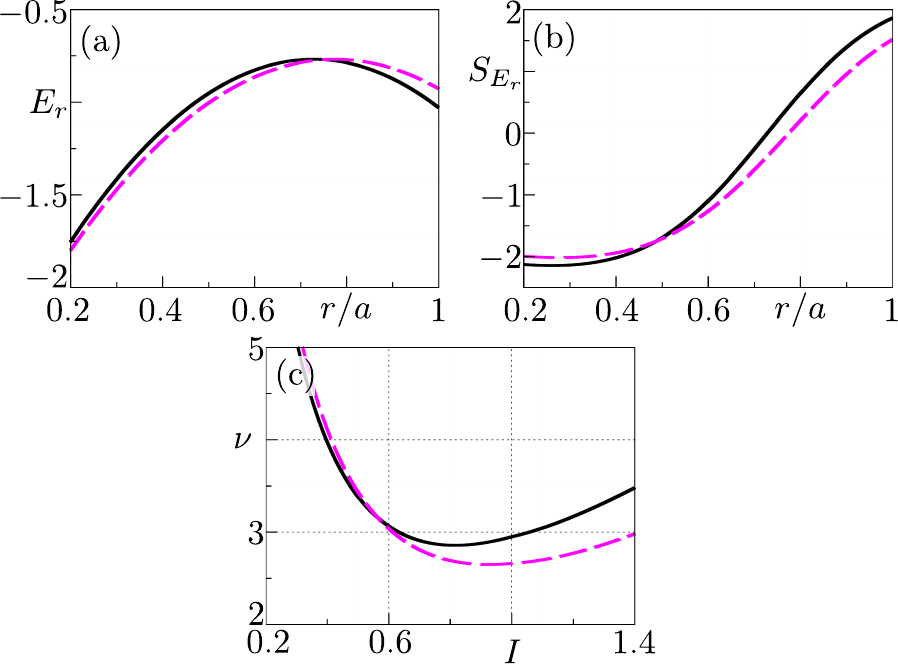}
    \caption{Radial profiles of (a) the equilibrium electric field, $E_r(r)$, (b) the $E_r$-shear, $S_{E_r}(r)$\textcolor{black}{,} and (c) the unperturbed Hamiltonian frequency, $\nu(I)$. For $I_{\mathrm{s}} = 0.53$ (\textcolor{black}{solid} black) and $I_{\mathrm{s}} = 0.60$ (\textcolor{black}{dashed} magenta). Note that now $n=2$ is nonresonant.}
    \label{fig:Nonmonotonic_Profiles_Er}
\end{figure}

We analyzed two different values of $I_{\mathrm{s}} = k^2$, $0.53$ and $0.60$, for which we present their $E_r(r)$, $S_{E_r}(r)$ and $\nu(I)$ profiles in Figure \ref{fig:Nonmonotonic_Profiles_Er}. In panels (a) and (b), we show that both profiles have a shearless point and, therefore, they are shear-reversed. This implies a nonmonotonic frequency profile $\nu(I)$, see panel (c) of the figure, and consequently a violation of the twist \mbox{condition \eqref{eq:twist.condition}}. On using these profiles, we can expect a shearless transport barrier to appear at $I_{\mathrm{STB}}$, for the unperturbed Hamiltonian, where $\left\lvert\partial \nu(I)/\partial I\right\lvert_{I_{\mathrm{STB}}}=0$. Meanwhile, for the perturbed system, the appearance of the shearless barrier depends on the values of $\phi_n$, as other works have shown \textcolor{black}{\cite{marcus2019,osorio2021}}. Also, note that now $n=2$ is nonresonant.               

For the perturbed Hamiltonian system, in order to find the shearless \textcolor{black}{curve}, \textcolor{black}{when} it exists,  it is necessary to calculate the rotation number profile, $\Omega(\psi_0,I_0)$. For this, we adopt a method based on a weighted Birkhoff average, which has been demonstrated to be superconvergent for calculating the rotation number of quasiperiodic orbits \cite{das2018}.\par 

Basically, \textcolor{black}{one uses} a function $g(\tau)$, that converges to zero with infinite smoothness at $\tau(0)=0$ and \mbox{$\tau(N)=1$}, with $\tau(j) = j/N$, to weight much less the beginning and the end of the orbit $\Sigma$ than the terms for which \mbox{$\tau \sim 0.50$}. \textcolor{black}{This function is also called bump.} In this work, we consider 
\begin{equation}
    g(\tau)= \left\{ \begin{array}{lcc}
			\exp\left(-\dfrac{1}{\tau^p(1-\tau)^p}\right), & \mathrm{for} & \tau \in (0,1), \cr\cr
			0, & \mathrm{for} & \tau \notin (0,1),
		\end{array}
		\right.
\end{equation}
with $p=1$, that provides satisfactory results \cite{das2017,sander2020}. \textcolor{black}{Other functions can be used, for example, for $p>1$ or even $g(\tau) = \sin^2(\pi\tau)$\textcolor{black}{; however, the latter one fails to be infinitely smooth at $\tau = 0$ and $\tau = 1$ \cite{das2017}}.}  

The rotation number, $\Omega$, of a regular orbit is independent of the choice of $\mathbf{P}_0$, as long as $\mathbf{P}_0 \in \Sigma$. However, the rotation number radial profile does depend on $(\psi_0, I)$. Different choices of $\psi_0$ will give different radial profiles, according to the distribution of the orbits on phase space. So, fixing $\psi_0$, the rotation number radial profile can be calculated as
\begin{subequations}\label{eq:rot}
    \begin{equation}
    \Omega(\psi_0,I) = \dfrac{1}{2\pi G_N}\sum_{j=0}^{N-1}g\left(\tau(j)\right)(\psi_{j+1}-\psi_j), 
\end{equation}
\begin{equation}
    G_N = \sum_{j=0}^{N}g(\tau(j)).
\end{equation}
\end{subequations}

\textcolor{black}{Note that for $g(\tau)=1$ the equation above returns to the usual definition of rotation number \cite{meiss1992}; however, the convergence is slower because assumes equal weights at any time of the orbit.}

\begin{figure}[htb]
    \centering
    \includegraphics[width = 0.45\textwidth]{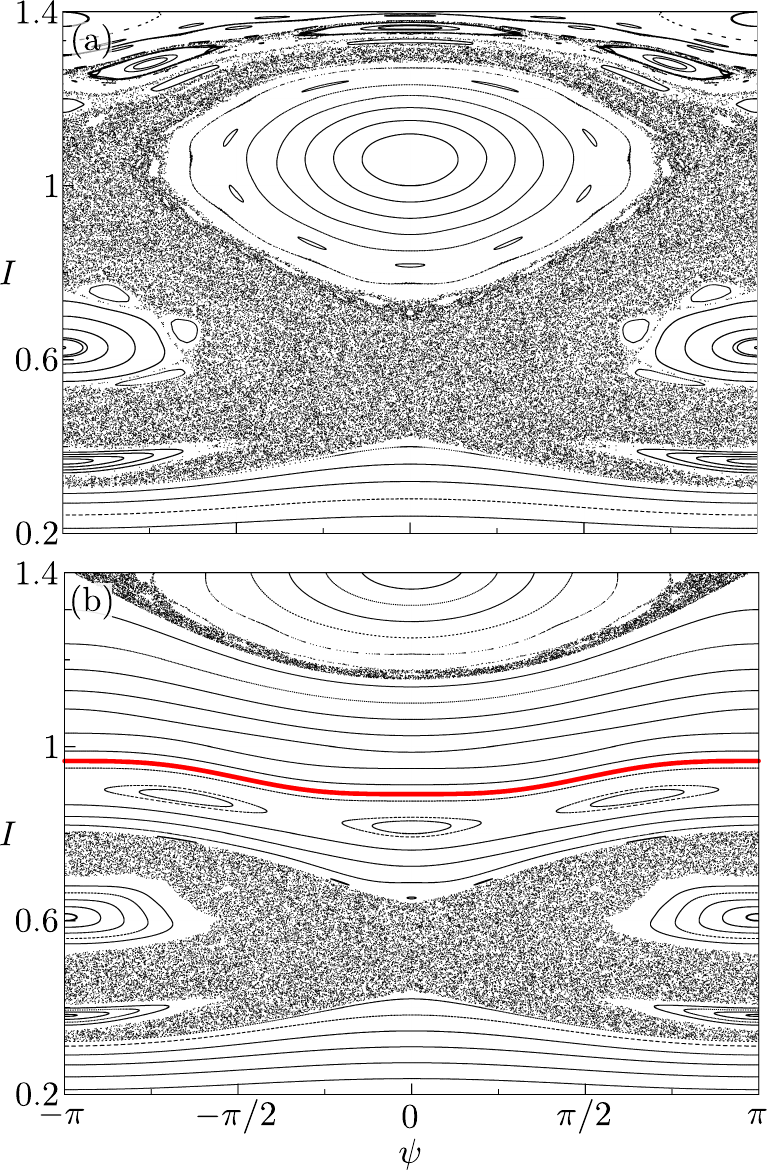}
    \caption{Poincar\'e sections for (a) $I_\mathrm{s} = 0.53$ and \mbox{(b) $I_\mathrm{s} = 0.60$}. A shearless curve is highlighted in red.}
    \label{fig:Phase_space_Er(Nonmonotonic)}
\end{figure}

\begin{figure}[htb]
    \centering
    \includegraphics[width = 0.47\textwidth]{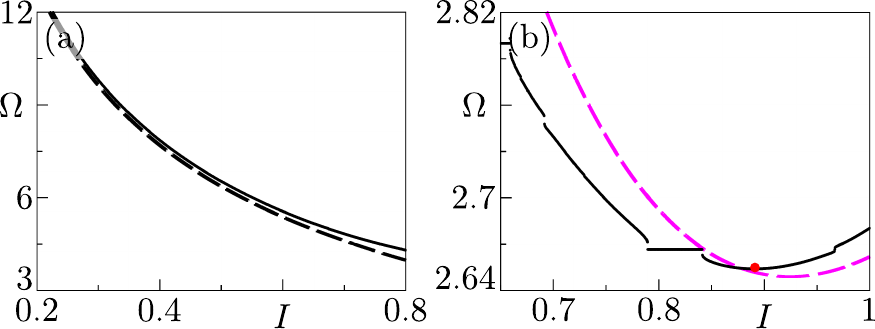}
    \caption{Rotation number profiles for (a) the unperturbed (dashed) and perturbed (\textcolor{black}{solid}) Hamiltonians associated with the linear radial electric field scenario, and rotation number profiles for (b) the unperturbed (\textcolor{black}{dashed} magenta) and perturbed (black) Hamiltonians associated with the nonmonotonic radial electric field.}
    \label{fig:RotNum_profiles2}
\end{figure}

From Figure \ref{fig:Phase_space_Er(Nonmonotonic)}, we see that, when $I_{\mathrm{s}}=0.53$, no resistance to the chaotic transport appears, most of the orbits escape outside the plasma. Nonetheless, by modifying the electric field, by taking $I_{\mathrm{s}} = 0.60$, a great suppressing of the chaotic transport occurs and, moreover, a shearless curve can be found close to the plasma edge, so confining most of the trajectories. The associated shearless curve was identified through the rotation number profile shown in Figure \ref{fig:RotNum_profiles2}(b), estimating the value $I_{\mathrm{STB}}$ for which $\left(\partial\Omega(\psi_0, I)/\partial I\right)_{I_{\mathrm{STB}}}=0$. The initial condition $\mathbf{P}_{\mathrm{STB}} = (\psi_0,I_{\mathrm{STB}})$ generates in phase space the invariant torus related to the barrier. 

Note that the $\Omega(\psi_0,I)$ profile for the linear $E_r(r)$ case does not have any shearless point, i.e., \textcolor{black}{it has} no shearless curve, panel (a) of the figure. As we explained before, the twist condition is not broken anywhere for that system. In fact,  for both the linear and the parabolic $E_r(r)$, rotation number profiles deviate just a little from the frequency profiles of the unperturbed Hamiltonian former cases.

\subsection{\textcolor{black}{Nonmonotonic magnetic field}}

{\color{black}
Besides the influence of the radial electric field profile, the equilibrium magnetic field profile can also change the transport in Horton's model. In the previous \textcolor{black}{sub}section, we applied a monotonic safety factor profile, together with a \textcolor{black}{monotonic and a} nonmonotonic electric field profile, Figure\textcolor{black}{s} \ref{fig:Monotonic_Profiles_Er} \textcolor{black}{and \ref{fig:Nonmonotonic_Profiles_Er}}. \textcolor{black}{The nonmonotonic} configuration results in the onset of a shearless curve in phase space, which can be broken up by the variation of the electric field profile, Figure \ref{fig:Phase_space_Er(Nonmonotonic)}. In this section, we will show that by modifying the safety factor with a nonmonotonic profile, the frequency profile of the system is changed and, consequently, its transport.

The nonmonotonic safety factor profile used in this paper is given by
\begin{equation}
    q(r) = q_\mathrm{m} + q_\mathrm{m}''(r - r_\mathrm{m})^2,
\end{equation}
where
\begin{equation}
    q_\mathrm{m}''(q_a) = \dfrac{q_a - q_\mathrm{m}}{(1-r_\mathrm{m})^2},
\end{equation}
and $r_\mathrm{m}$ is the radial location of the safety factor profile minimum, $q_\mathrm{m}=q(r_\mathrm{m})$. The safety factor at the plasma edge, $q_a$, is the control parameter chosen to analyze the influence of the magnetic field profile. It is related to the plasma current, an easily obtained experimental measure \cite{wesson2011}.

We choose two values of the safety factor at the plasma edge to analyze the Poincaré section, $q_a=5.00$ and \mbox{$q_a=3.65$}. The corresponding $q(r)$ and frequency profiles are shown in Figure \ref{fig:Monmonotonic_Profiles_q}. The other profile parameters \textcolor{black}{are} fixed \textcolor{black}{at} $r_\mathrm{m}=0.4\textcolor{black}{0}$ and $q_\mathrm{m}=2\textcolor{black}{.00}$. \textcolor{black}{For the electric field radial profile we use $I_{\mathrm{s}} = 0.5$.} 

The same resonant modes, $n=3$ and $4$ are found in both nonmonotonic safety factor configurations. They also have a shearless point at $I\textcolor{black}{_{\mathrm{STB}}}\approx0.8$. For $q_a=3.65$, the \textcolor{black}{two} resonances \textcolor{black}{associated with $n=3$} are close to each \textcolor{black}{other}, as seen in the \textcolor{black}{inset} \textcolor{black}{of} Figure \ref{fig:Monmonotonic_Profiles_q}(b).

\begin{figure}[htb]
    \centering
    \includegraphics[width=.47\textwidth]{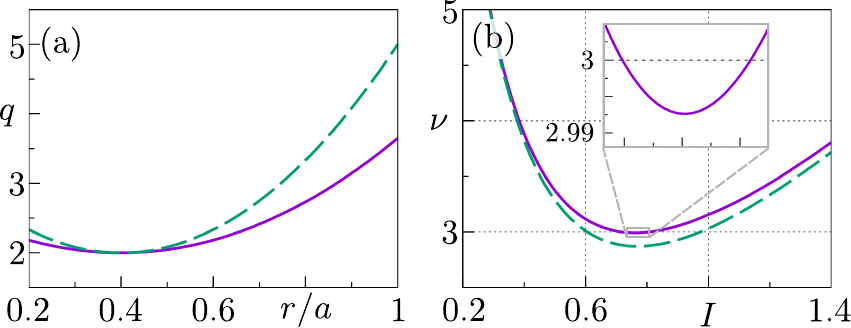}
    \caption{(a) Nonmonotonic safety factor profile and (b) corresponding resonance condition for $q_a=3.65$ (\textcolor{black}{solid} purple) and $q_a=5.00$ (\textcolor{black}{dashed} green).}
    \label{fig:Monmonotonic_Profiles_q}
\end{figure}

Figure \ref{fig:Phase_Space_q(Nonmonotonic)} shows the Poincaré sections for \mbox{$q_a=5.00$} and $q_a=3.65$. In the first one, the configuration of orbits indicates a large transport: the chaotic orbits spread in a large portion of phase space, crossing the plasma edge\textcolor{black}{, at} $I=1\textcolor{black}{.00}$, Figure \ref{fig:Monmonotonic_Profiles_q}(a). The larger islands are associated with the resonant mode $n=3$, localized at $I\approx 0.6$ and $I\approx 1.0$, Figure \ref{fig:Monmonotonic_Profiles_q}. The same islands are present for $q_a=3.65$, although their radial location\textcolor{black}{s} \textcolor{black}{are close to $I_{\mathrm{STB}}$}, Figure \ref{fig:Monmonotonic_Profiles_q}(b). For this $q_a$ value, the phase space presents a shearless curve, marked in red, that acts as a transport barrier, preventing internal chaotic orbits to be transported across the plasma edge.

\begin{figure}[htb]
    \centering
    \includegraphics[width = 0.45\textwidth]{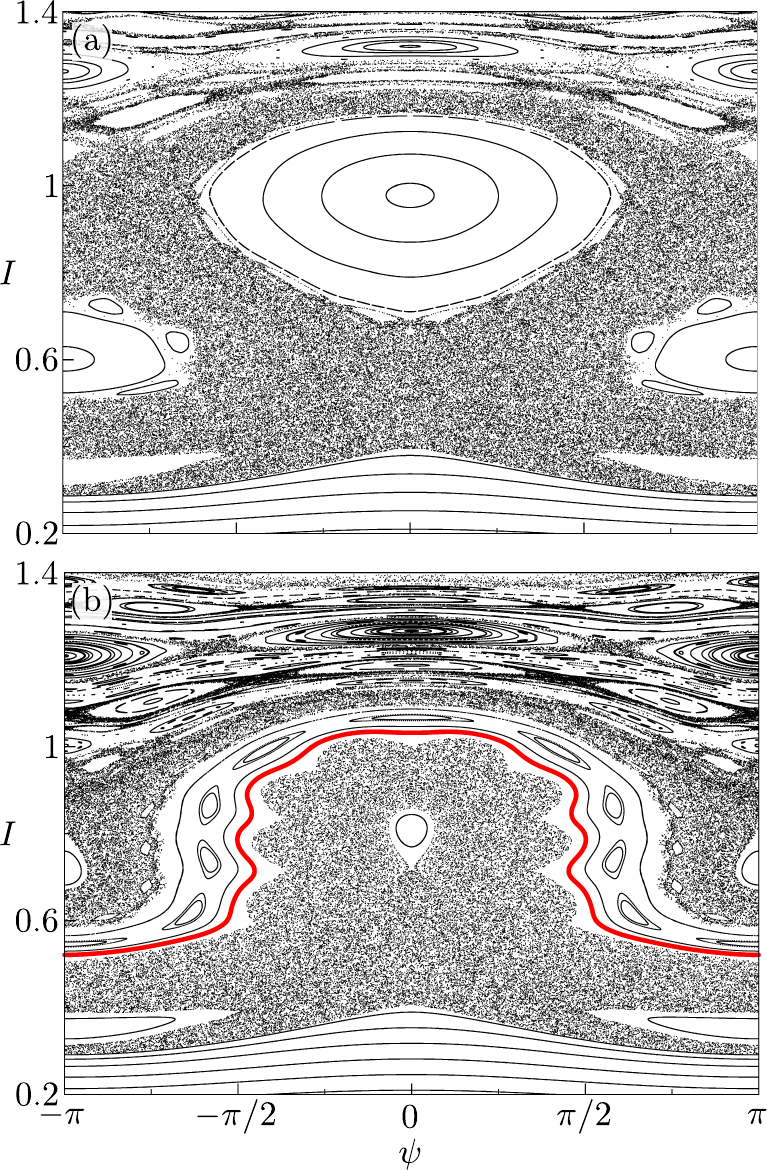}
    \caption{Poincar\'e section using nonmonotonic safety factor profile for (a) $q_a=5.0$ and (b) $q_a=3.65$. The shearless invariant curve (red) acts as a robust transport barrier in phase space.}
    \label{fig:Phase_Space_q(Nonmonotonic)}
\end{figure}

In summary, the nonmonotonic safety factor changes the frequency profile of the system, and consequently, its resonance conditions. The results in this \textcolor{black}{sub}section indicate a relation between the safety factor profile with the onset and breakup of the shearless curve, which is a barrier \textcolor{black}{to} transport in phase space. In the next section, we will show that the presence of a shearless curve in phase space systematically reduces the transport coefficient of the system, even when the fluctuating amplitude modes are increased.
}

\section{Transport coefficient analysis \label{sec:Transport_coefficient}}

\begin{figure*}
    \centering
    \includegraphics[width = \textwidth]{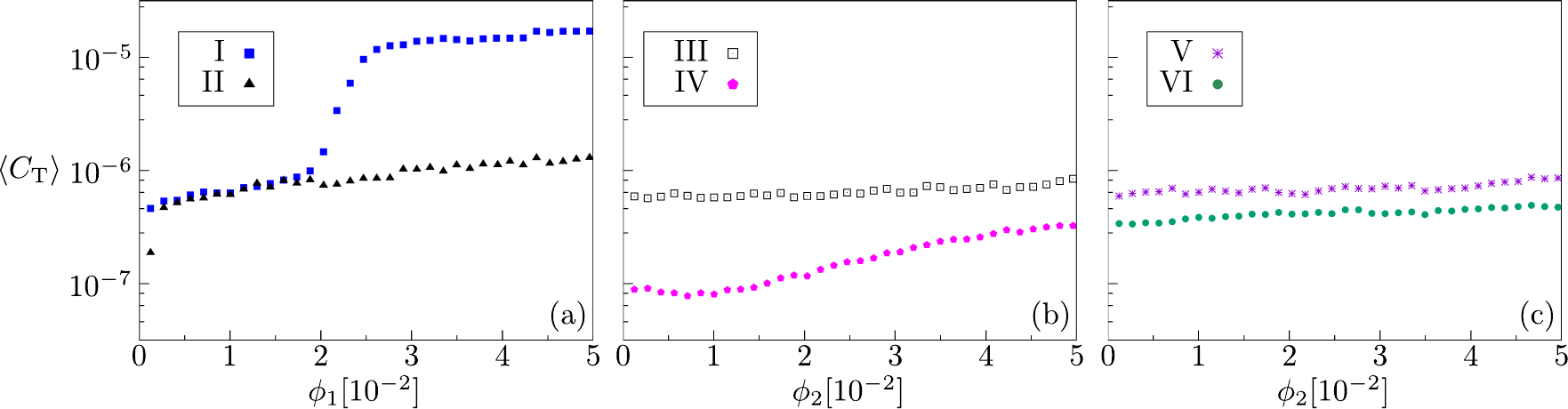}
    \caption{(a) $E_r$ constant (\textcolor{black}{filled blue squares}) and $E_r$ linear (\textcolor{black}{filled black triangles}). (b) $I_{\mathrm{s}}=0.53$ (\textcolor{black}{empty black squares}) and $I_{\mathrm{s}}=0.60$ (\textcolor{black}{filled magenta pentagons}). (c) $q_a=5.00$ (\textcolor{black}{filled green circles}) and $q_a=3.65$ (\textcolor{black}{purple crosses}).}
    \label{fig:Difusion}
\end{figure*}

{\color{black}
In this section, we analyze the chaotic transport dependence on the amplitude of the fluctuating modes $\phi_n$ for each one of the plasma profile configurations used in Figures \ref{fig:Phase_Space_Er(Monotonic)}, \ref{fig:Phase_space_Er(Nonmonotonic)} and \ref{fig:Phase_Space_q(Nonmonotonic)}. Those profiles are categorized into six cases, as shown in Table \ref{tab:configurations}.

\begin{table}[htb]
    \centering
    \begin{tabular}{|c|c|c|}
    \hline
    Case & $E_r$ profile & $q$ profile \\
    \hline
    I & constant & monoto. \\
    \hline
    II & linear & monot. \\
    \hline
    III & nonmonot. $k=0.53$ & monot. \\
    \hline
    IV & nonmonot. $k=0.60$ & monot. \\
    \hline
    V & nonmonot. $k=0.50$ & nonmonot. $q_a=5.00$ \\
    \hline
    VI & nonmonot. $k=0.50$ & nonmonot. $q_a=3.65$ \\
    \hline
    \end{tabular}
    \caption{Plasma configurations for the equilibrium electric and magnetic fields radial profiles considered in this article.}
    \label{tab:configurations}
\end{table}
To quantify this transport, we calculate the transport coefficient $\langle C_T \rangle$, which is the average of the running transport coefficient $C_T(t)$, a quantification of the \textcolor{black}{radial} transport in phase space. More precisely, given an ensemble of $\mathcal{N}$ orbits,  the running transport coefficient is proportional to the mean squared deviation of the action variables $I(t)$ with \textcolor{black}{respect} to the initial conditions $I(0)$\textcolor{black}{;} \textcolor{black}{it} is given by
\begin{equation}
    C_T(t) = \dfrac{1}{2t\mathcal{N}} \sum_{\textcolor{black}{\ell}=1}^{\mathcal{N}}\left[ I_{}^{\textcolor{black}{(\ell)}}(t) - I_{}^{\textcolor{black}{(\ell)}}(0) \right]^2,
\end{equation}
where $I^{\textcolor{black}{(\ell)}} (t)$ is the action of the $\textcolor{black}{\ell}$-th particle of the ensemble, at the time $t$. The mean transport coefficient is given by the time average of $C_T(t)$, i.e.,
\begin{equation}
    \langle C_T \rangle = \dfrac{1}{t_{\mathrm{f}}-t_{\mathrm{i}}} \sum_{t_j=t_{\mathrm{i}}}^{t_{\mathrm{f}}} (C_T(t_j))
\end{equation}
where $t_{\mathrm{i}}$ is the time at which the convergence of $C_T(t)$ is observed to set in.

In the ensemble, some orbits will be chaotic, but others are regular (invariant curves or islands). The chaotic orbits will contribute to increasing $\langle C_T \rangle$, and the regular ones tend to decrease it. As the coefficient is an average over a large ensemble, higher values of $\langle C_T \rangle$ indicate large portions of chaotic orbits in the phase space, \textcolor{black}{which} have a large transport in the action coordinate. On the other hand, small values of $\langle C_T \rangle$ indicate a small portion of chaotic orbits in phase space\textcolor{black}{, which} have a small transport in action coordinate. For numerical purposes, we compute\textcolor{black}{d} the running transport coefficient for an ensemble of $\mathcal{N}=10^3$ \textcolor{black}{ramdonly chosen} orbits, with \textcolor{black}{action initial conditions chosen such that $I^{(\ell)}(0)\in[0.2,1]$}, \textcolor{black}{which were} integrated until \mbox{$t_{\mathrm{f}}=(2\pi/\omega_0)\times 10^5$}, with a transient of $\textcolor{black}{t_{\mathrm{i}}=0.3\,t_{\mathrm{f}}}$.

We computed the transport coefficient $\textcolor{black}{\langle} C_T\textcolor{black}{\rangle}$ varying the amplitude of the higher \textcolor{black}{associated} nonresonant mode, which means, for the configuration\textcolor{black}{s} of Figure \ref{fig:Monotonic_Profiles_Er} (Cases I and II)\textcolor{black}{,} the nonresonant mode $n=1$, and for the plasma profiles of Figures \ref{fig:Nonmonotonic_Profiles_Er} (Cases III and IV) and \ref{fig:Monmonotonic_Profiles_q} (Cases V and VI)\textcolor{black}{, the nonresonant mode $n=2$}. The associated nonresonant modes amplitude\textcolor{black}{s} \textcolor{black}{were} varied from $0$ to $5\times 10^{-2}$\textcolor{black}{,} and the obtained results are present\textcolor{black}{ed} in Figure \ref{fig:Difusion}.

Both configurations with monotonic radial electric fields have the higher transport coefficient values among all \textcolor{black}{the} configurations \textcolor{black}{considered}. The results in Figure \ref{fig:Difusion}(a) indicate a similar transport coefficient until \mbox{$\phi_1\approx 2\times 10^{-2}$}, and for \textcolor{black}{higher} values, the transport for Case I is substantially higher \textcolor{black}{than} {\textcolor{black}{for} Case II. This result indicates that the radial electric field shear $S_{E_r}$ provides suppression of chaotic orbits in phase space, as identified in Figure \ref{fig:Phase_Space_Er(Monotonic)}.

Applying nonmonotonic radial electric field profiles, \textcolor{black}{we can reduce} the transport coefficient in comparison with the monotonic profiles \textcolor{black}{scenarios}, Figure \ref{fig:Difusion}(b). Considering Cases III and IV, \textcolor{black}{we found that} the \textcolor{black}{latter} has a smaller coefficient, \textcolor{black}{therefore,} the shearless curve present\textcolor{black}{ed} in Figure \ref{fig:Phase_space_Er(Nonmonotonic)}(b) provides a robust barrier to the transport in phase space. \textcolor{black}{Futhermore}, the transport coefficient of Case III is similar to Case II, indicating that the shearless curve is not present in the phase space of III, for any value of $\phi_2$.

Finally, the influence of the safety factor profile in the transport coefficient is present\textcolor{black}{ed} in Figure \ref{fig:Difusion}(c). For Case V, in comparison to Case III, the addition of a nonmonotonic $q$ profile does not change the transport coefficient. However, for Case VI, the transport coefficient is reduced, due to the presence of the shearless curve in phase space, Figure \ref{fig:Phase_Space_q(Nonmonotonic)}(b). In summary, the nonmonotonic safety factor profile can reduce the transport in phase space, and its profile influences the onset and breakup of the shearless curve.
}

\section{Conclusions \label{sec:conclusions}}
{\color{black}
In this work, we applied a transport model to describe the trajectory of test particles at the tokamak plasma edge subject to the $\mathbf{E}\times\mathbf{B}$ drift motion, associated with anomalous transport. Numerical simulations of the corresponding Hamiltonian system were done to obtain the trajectory of the test particle, plotted at a Poincaré section. Those orbits can be regular or chaotic, associated with particle transport.

We considered different configurations of plasma electric and magnetic field profiles. Two monotonic electric field profiles were considered, a constant and a linear one. The Poincaré sections obtained indicate low confinement of orbits \textcolor{black}{for} the constant electric field configuration. In addition, the linear profile, which has an electric shear, presented a suppression of the chaotic orbits.

Considering nonmonotonic electric field profiles, with reversed shear, chaos suppression is also present. Moreover, in such systems with nonmonotonic profiles, the phase space can present a shearless curve, that acts as a robust transport barrier in phase space, preventing the chaotic orbits \textcolor{black}{from leaving} the plasma. However, this transport barrier can be broken \textcolor{black}{up} by varying the plasma \textcolor{black}{equilibrium electric field} profile.

In the last plasma configuration, both \textcolor{black}{the} electric and magnetic field profiles \textcolor{black}{were} considered nonmonotonic. This configuration also presented the shearless transport barrier. Nonetheless, the results obtained indicate that, by variations in the safety factor profiles, the shearless curve could also be broken \textcolor{black}{up}.

Finally, we made an analysis of the transport coefficient in each plasma configuration presented in this paper. The transport coefficient was numerically obtained by calculating the squared mean deviation of the particle\textcolor{black}{s} orbit\textcolor{black}{s}, as a function of a perturbation mode amplitude. The results indicate low confinement for a constant electric field. The addition of sheared electric field profiles, such as in \textcolor{black}{the} linear or parabolic cases, reduced the transport coefficients. \textcolor{black}{On i}ntroducing nonmonotonic profiles of \textcolor{black}{the} electric and magnetic fields\textcolor{black}{,} shearless transport barriers appear \textcolor{black}{which} \textcolor{black}{reduce} even more the \textcolor{black}{radial} transport coefficient.

In summary, the results in this paper indicate that by using Horton's mode\textcolor{black}{l} to study the anomalous transport of particles at the plasma edge, sheared plasma profiles tend to reduce the plasma transport, as predicted by the $\mathbf{E}\times\mathbf{B}$ shear stabilization models. They  predict a reduction in transport where the electric shear is high. \textcolor{black}{Nevertheless, the effect of shearless transport barriers, associated with regions of null shear, present in nonmonotonic profiles, reduces the transport}.

\begin{acknowledgments}
The authors thank the financial support from the Brazilian Federal Agencies (CNPq), grants 304616/2021-4 and 302665/2017-0, the S{\~a}o Paulo Research Foundation (FAPESP, Brazil) under grants 2018/03211-6, 2020/01399-8, 2022/04251-7, 2022/08699-2 and 2022/05667-2, the Coordena\c{c}\~{a}o de Aperfei\c{c}oamento de Pessoal de N{\'i}vel Superior (CAPES) under Grant No. 88881.143103/2017-01, and the  Comit{\'e} Fran\c{c}ais d'Evaluation de la Coop{\'e}ration Universitaire et Scientifique avec le Br{\'e}sil (COFECUB) under Grant No. 40273QA-Ph908/18.\par

The Centre de Calcul Intensif d'Aix-Marseille is acknowledged for granting access to its high performance computing resources.\par
L.A. Osorio-Quiroga thanks Cristel Chandre for advising him on the rotation number calculation. 
\end{acknowledgments}

\appendix
\section{\textcolor{black}{Normalization of the model equations}}\label{sec:appendix}
{\color{black}
In this appendix, we outline the \textcolor{black}{normalization} of the variables and parameters of the Hamiltonian system used in this paper. The characteristic scales of the system are chosen as \mbox{$a_c=a$ (length)}, $E_c=\lvert E_r(r=a) \rvert$ (electric field) and $B_0=B_c$ (magnetic field). That is, the normalized quantities, denoted with a prime, are
\begin{equation}
    a' = \dfrac{a}{a_c}, \ E_r'=\dfrac{E_r}{E_c}, \ B'=\dfrac{B}{B_c}.
\end{equation}
The remaining parameters are chosen according to
\begin{subequations}
    \begin{align}
    \phi_n' &= \dfrac{\phi_n}{aE_c}, \ \ v_{\parallel}' = \dfrac{B_c}{E_c}v_{\parallel}\\
    t' &= \dfrac{E_c}{aB_c}t, \ \ \omega_0' = \dfrac{a_cB_c}{E_c}\omega_0.
\end{align}
\end{subequations}

\textcolor{black}{Finally}, the numerical values of the parameters are: $\omega_0=5.7$, $v_{\parallel} = 2.0$, and \mbox{$(\phi_2,\phi_3,\phi_4) = (1.95,3.66,2.08)\times 10^{-3}.$}
}

\section*{References}

\bibliography{sn-bibliography}

%merlin.mbs aipnum4-1.bst 2010-07-25 4.21a (PWD, AO, DPC) hacked
%Control: key (0)
%Control: author (8) initials jnrlst
%Control: editor formatted (1) identically to author
%Control: production of article title (0) allowed
%Control: page (1) range
%Control: year (1) truncated
%Control: production of eprint (0) enabled
\begin{thebibliography}{38}%
\makeatletter
\providecommand \@ifxundefined [1]{%
 \@ifx{#1\undefined}
}%
\providecommand \@ifnum [1]{%
 \ifnum #1\expandafter \@firstoftwo
 \else \expandafter \@secondoftwo
 \fi
}%
\providecommand \@ifx [1]{%
 \ifx #1\expandafter \@firstoftwo
 \else \expandafter \@secondoftwo
 \fi
}%
\providecommand \natexlab [1]{#1}%
\providecommand \enquote  [1]{``#1''}%
\providecommand \bibnamefont  [1]{#1}%
\providecommand \bibfnamefont [1]{#1}%
\providecommand \citenamefont [1]{#1}%
\providecommand \href@noop [0]{\@secondoftwo}%
\providecommand \href [0]{\begingroup \@sanitize@url \@href}%
\providecommand \@href[1]{\@@startlink{#1}\@@href}%
\providecommand \@@href[1]{\endgroup#1\@@endlink}%
\providecommand \@sanitize@url [0]{\catcode `\\12\catcode `\$12\catcode
  `\&12\catcode `\#12\catcode `\^12\catcode `\_12\catcode `\%12\relax}%
\providecommand \@@startlink[1]{}%
\providecommand \@@endlink[0]{}%
\providecommand \url  [0]{\begingroup\@sanitize@url \@url }%
\providecommand \@url [1]{\endgroup\@href {#1}{\urlprefix }}%
\providecommand \urlprefix  [0]{URL }%
\providecommand \Eprint [0]{\href }%
\providecommand \doibase [0]{http://dx.doi.org/}%
\providecommand \selectlanguage [0]{\@gobble}%
\providecommand \bibinfo  [0]{\@secondoftwo}%
\providecommand \bibfield  [0]{\@secondoftwo}%
\providecommand \translation [1]{[#1]}%
\providecommand \BibitemOpen [0]{}%
\providecommand \bibitemStop [0]{}%
\providecommand \bibitemNoStop [0]{.\EOS\space}%
\providecommand \EOS [0]{\spacefactor3000\relax}%
\providecommand \BibitemShut  [1]{\csname bibitem#1\endcsname}%
\let\auto@bib@innerbib\@empty
%</preamble>
\bibitem [{\citenamefont {Cowley}(2016)}]{cowley2016}%
  \BibitemOpen
  \bibfield  {author} {\bibinfo {author} {\bibfnamefont {S.~C.}\ \bibnamefont
  {Cowley}},\ }\bibfield  {title} {\enquote {\bibinfo {title} {The quest for
  fusion power},}\ }\href@noop {} {\bibfield  {journal} {\bibinfo  {journal}
  {Nature physics}\ }\textbf {\bibinfo {volume} {12}},\ \bibinfo {pages} {384}
  (\bibinfo {year} {2016})}\BibitemShut {NoStop}%
\bibitem [{\citenamefont {Horton}(2018)}]{horton2018}%
  \BibitemOpen
  \bibfield  {author} {\bibinfo {author} {\bibfnamefont {W.}~\bibnamefont
  {Horton}},\ }\href@noop {} {\enquote {\bibinfo {title} {Turbulent transport
  in magnetized plasmas},}\ } (\bibinfo {year} {2018})\BibitemShut {NoStop}%
\bibitem [{\citenamefont {Wolf}(2002)}]{wolf2002}%
  \BibitemOpen
  \bibfield  {author} {\bibinfo {author} {\bibfnamefont {R.}~\bibnamefont
  {Wolf}},\ }\bibfield  {title} {\enquote {\bibinfo {title} {Internal transport
  barriers in tokamak plasmas},}\ }\href@noop {} {\bibfield  {journal}
  {\bibinfo  {journal} {Plasma Physics and Controlled Fusion}\ }\textbf
  {\bibinfo {volume} {45}},\ \bibinfo {pages} {R1} (\bibinfo {year}
  {2002})}\BibitemShut {NoStop}%
\bibitem [{\citenamefont {Cavedon}\ \emph {et~al.}(2016)\citenamefont
  {Cavedon}, \citenamefont {P{\"u}tterich}, \citenamefont {Viezzer},
  \citenamefont {Birkenmeier}, \citenamefont {Happel}, \citenamefont {Laggner},
  \citenamefont {Manz}, \citenamefont {Ryter}, \citenamefont {Stroth},
  \citenamefont {{ASDEX Upgrade Team}} \emph {et~al.}}]{cavedon2016}%
  \BibitemOpen
  \bibfield  {author} {\bibinfo {author} {\bibfnamefont {M.}~\bibnamefont
  {Cavedon}}, \bibinfo {author} {\bibfnamefont {T.}~\bibnamefont
  {P{\"u}tterich}}, \bibinfo {author} {\bibfnamefont {E.}~\bibnamefont
  {Viezzer}}, \bibinfo {author} {\bibfnamefont {G.}~\bibnamefont
  {Birkenmeier}}, \bibinfo {author} {\bibfnamefont {T.}~\bibnamefont {Happel}},
  \bibinfo {author} {\bibfnamefont {F.}~\bibnamefont {Laggner}}, \bibinfo
  {author} {\bibfnamefont {P.}~\bibnamefont {Manz}}, \bibinfo {author}
  {\bibfnamefont {F.}~\bibnamefont {Ryter}}, \bibinfo {author} {\bibfnamefont
  {U.}~\bibnamefont {Stroth}}, \bibinfo {author} {\bibnamefont {{ASDEX Upgrade
  Team}}},  \emph {et~al.},\ }\bibfield  {title} {\enquote {\bibinfo {title}
  {Interplay between turbulence, neoclassical and zonal flows during the
  transition from low to high confinement mode at {ASDEX} upgrade},}\
  }\href@noop {} {\bibfield  {journal} {\bibinfo  {journal} {Nuclear Fusion}\
  }\textbf {\bibinfo {volume} {57}},\ \bibinfo {pages} {014002} (\bibinfo
  {year} {2016})}\BibitemShut {NoStop}%
\bibitem [{\citenamefont {Burrell}(1997)}]{burrell1997}%
  \BibitemOpen
  \bibfield  {author} {\bibinfo {author} {\bibfnamefont {K.~H.}\ \bibnamefont
  {Burrell}},\ }\bibfield  {title} {\enquote {\bibinfo {title} {Effects of
  {E}×{B} velocity shear and magnetic shear on turbulence and transport in
  magnetic confinement devices},}\ }\href@noop {} {\bibfield  {journal}
  {\bibinfo  {journal} {Phys. Plasmas}\ }\textbf {\bibinfo {volume} {4}},\
  \bibinfo {pages} {1499} (\bibinfo {year} {1997})}\BibitemShut {NoStop}%
\bibitem [{\citenamefont {Connor}\ \emph {et~al.}(2004)\citenamefont {Connor},
  \citenamefont {Fukuda}, \citenamefont {Garbet}, \citenamefont {Gormezano},
  \citenamefont {Mukhovatov},\ and\ \citenamefont {Wakatani}}]{connor2004}%
  \BibitemOpen
  \bibfield  {author} {\bibinfo {author} {\bibfnamefont {J.~W.}\ \bibnamefont
  {Connor}}, \bibinfo {author} {\bibfnamefont {T.}~\bibnamefont {Fukuda}},
  \bibinfo {author} {\bibfnamefont {X.}~\bibnamefont {Garbet}}, \bibinfo
  {author} {\bibfnamefont {C.}~\bibnamefont {Gormezano}}, \bibinfo {author}
  {\bibfnamefont {V.}~\bibnamefont {Mukhovatov}}, \ and\ \bibinfo {author}
  {\bibfnamefont {M.}~\bibnamefont {Wakatani}},\ }\bibfield  {title} {\enquote
  {\bibinfo {title} {A review of internal transport barrier physics for
  steady-state operation of tokamaks},}\ }\href@noop {} {\bibfield  {journal}
  {\bibinfo  {journal} {Nucl. Fusion}\ }\textbf {\bibinfo {volume} {44}},\
  \bibinfo {pages} {R1} (\bibinfo {year} {2004})}\BibitemShut {NoStop}%
\bibitem [{\citenamefont {Garbet}\ \emph {et~al.}(2004)\citenamefont {Garbet},
  \citenamefont {Mantica}, \citenamefont {Angioni}, \citenamefont {Asp},
  \citenamefont {Baranov}, \citenamefont {Bourdelle}, \citenamefont {Budny},
  \citenamefont {Crisanti}, \citenamefont {Cordey}, \citenamefont {Garzotti}
  \emph {et~al.}}]{garbet2004}%
  \BibitemOpen
  \bibfield  {author} {\bibinfo {author} {\bibfnamefont {X.}~\bibnamefont
  {Garbet}}, \bibinfo {author} {\bibfnamefont {P.}~\bibnamefont {Mantica}},
  \bibinfo {author} {\bibfnamefont {C.}~\bibnamefont {Angioni}}, \bibinfo
  {author} {\bibfnamefont {E.}~\bibnamefont {Asp}}, \bibinfo {author}
  {\bibfnamefont {Y.}~\bibnamefont {Baranov}}, \bibinfo {author} {\bibfnamefont
  {C.}~\bibnamefont {Bourdelle}}, \bibinfo {author} {\bibfnamefont
  {R.}~\bibnamefont {Budny}}, \bibinfo {author} {\bibfnamefont
  {F.}~\bibnamefont {Crisanti}}, \bibinfo {author} {\bibfnamefont
  {G.}~\bibnamefont {Cordey}}, \bibinfo {author} {\bibfnamefont
  {L.}~\bibnamefont {Garzotti}},  \emph {et~al.},\ }\bibfield  {title}
  {\enquote {\bibinfo {title} {Physics of transport in tokamaks},}\ }\href@noop
  {} {\bibfield  {journal} {\bibinfo  {journal} {Plasma Physics and Controlled
  Fusion}\ }\textbf {\bibinfo {volume} {46}},\ \bibinfo {pages} {B557}
  (\bibinfo {year} {2004})}\BibitemShut {NoStop}%
\bibitem [{\citenamefont {Moyer}\ \emph {et~al.}(1995)\citenamefont {Moyer},
  \citenamefont {Burrell}, \citenamefont {Carlstrom}, \citenamefont {Coda},
  \citenamefont {Conn}, \citenamefont {Doyle}, \citenamefont {Gohil},
  \citenamefont {Groebner}, \citenamefont {Kim}, \citenamefont {Lehmer} \emph
  {et~al.}}]{moyer1995}%
  \BibitemOpen
  \bibfield  {author} {\bibinfo {author} {\bibfnamefont {R.}~\bibnamefont
  {Moyer}}, \bibinfo {author} {\bibfnamefont {K.}~\bibnamefont {Burrell}},
  \bibinfo {author} {\bibfnamefont {T.}~\bibnamefont {Carlstrom}}, \bibinfo
  {author} {\bibfnamefont {S.}~\bibnamefont {Coda}}, \bibinfo {author}
  {\bibfnamefont {R.}~\bibnamefont {Conn}}, \bibinfo {author} {\bibfnamefont
  {E.}~\bibnamefont {Doyle}}, \bibinfo {author} {\bibfnamefont
  {P.}~\bibnamefont {Gohil}}, \bibinfo {author} {\bibfnamefont
  {R.}~\bibnamefont {Groebner}}, \bibinfo {author} {\bibfnamefont
  {J.}~\bibnamefont {Kim}}, \bibinfo {author} {\bibfnamefont {R.}~\bibnamefont
  {Lehmer}},  \emph {et~al.},\ }\bibfield  {title} {\enquote {\bibinfo {title}
  {Beyond paradigm: {T}urbulence, transport, and the origin of the radial
  electric field in low to high confinement mode transitions in the {DIII-D}
  tokamak},}\ }\href@noop {} {\bibfield  {journal} {\bibinfo  {journal}
  {Physics of Plasmas}\ }\textbf {\bibinfo {volume} {2}},\ \bibinfo {pages}
  {2397} (\bibinfo {year} {1995})}\BibitemShut {NoStop}%
\bibitem [{\citenamefont {Kobayashi}\ \emph {et~al.}(2017)\citenamefont
  {Kobayashi}, \citenamefont {Itoh}, \citenamefont {Ido}, \citenamefont
  {Kamiya}, \citenamefont {Itoh}, \citenamefont {Miura}, \citenamefont
  {Nagashima}, \citenamefont {Fujisawa}, \citenamefont {Inagaki},\ and\
  \citenamefont {Ida}}]{kobayashi2017}%
  \BibitemOpen
  \bibfield  {author} {\bibinfo {author} {\bibfnamefont {T.}~\bibnamefont
  {Kobayashi}}, \bibinfo {author} {\bibfnamefont {K.}~\bibnamefont {Itoh}},
  \bibinfo {author} {\bibfnamefont {T.}~\bibnamefont {Ido}}, \bibinfo {author}
  {\bibfnamefont {K.}~\bibnamefont {Kamiya}}, \bibinfo {author} {\bibfnamefont
  {S.-I.}\ \bibnamefont {Itoh}}, \bibinfo {author} {\bibfnamefont
  {Y.}~\bibnamefont {Miura}}, \bibinfo {author} {\bibfnamefont
  {Y.}~\bibnamefont {Nagashima}}, \bibinfo {author} {\bibfnamefont
  {A.}~\bibnamefont {Fujisawa}}, \bibinfo {author} {\bibfnamefont
  {S.}~\bibnamefont {Inagaki}}, \ and\ \bibinfo {author} {\bibfnamefont
  {K.}~\bibnamefont {Ida}},\ }\bibfield  {title} {\enquote {\bibinfo {title}
  {Turbulent transport reduction induced by transition on radial electric field
  shear and curvature through amplitude and cross-phase in torus plasma},}\
  }\href@noop {} {\bibfield  {journal} {\bibinfo  {journal} {Scientific
  reports}\ }\textbf {\bibinfo {volume} {7}},\ \bibinfo {pages} {14971}
  (\bibinfo {year} {2017})}\BibitemShut {NoStop}%
\bibitem [{\citenamefont {Kamiya}, \citenamefont {Itoh},\ and\ \citenamefont
  {Itoh}(2016)}]{kamiya2016}%
  \BibitemOpen
  \bibfield  {author} {\bibinfo {author} {\bibfnamefont {K.}~\bibnamefont
  {Kamiya}}, \bibinfo {author} {\bibfnamefont {K.}~\bibnamefont {Itoh}}, \ and\
  \bibinfo {author} {\bibfnamefont {S.-I.}\ \bibnamefont {Itoh}},\ }\bibfield
  {title} {\enquote {\bibinfo {title} {Experimental validation of
  non-uniformity effect of the radial electric field on the edge transport
  barrier formation in {JT-60U H}-mode plasmas},}\ }\href@noop {} {\bibfield
  {journal} {\bibinfo  {journal} {Scientific reports}\ }\textbf {\bibinfo
  {volume} {6}},\ \bibinfo {pages} {30585} (\bibinfo {year}
  {2016})}\BibitemShut {NoStop}%
\bibitem [{\citenamefont {Caldas}\ \emph {et~al.}(2012)\citenamefont {Caldas},
  \citenamefont {Viana}, \citenamefont {Abud}, \citenamefont {Fonseca},
  \citenamefont {Guimar{\~a}es~Filho}, \citenamefont {Kroetz}, \citenamefont
  {Marcus}, \citenamefont {Schelin}, \citenamefont {Szezech}, \citenamefont
  {Toufen} \emph {et~al.}}]{caldas2012}%
  \BibitemOpen
  \bibfield  {author} {\bibinfo {author} {\bibfnamefont {I.~L.}\ \bibnamefont
  {Caldas}}, \bibinfo {author} {\bibfnamefont {R.~L.}\ \bibnamefont {Viana}},
  \bibinfo {author} {\bibfnamefont {C.~V.}\ \bibnamefont {Abud}}, \bibinfo
  {author} {\bibfnamefont {J.~C. D.~d.}\ \bibnamefont {Fonseca}}, \bibinfo
  {author} {\bibfnamefont {Z.~d.~O.}\ \bibnamefont {Guimar{\~a}es~Filho}},
  \bibinfo {author} {\bibfnamefont {T.}~\bibnamefont {Kroetz}}, \bibinfo
  {author} {\bibfnamefont {F.~A.}\ \bibnamefont {Marcus}}, \bibinfo {author}
  {\bibfnamefont {A.~B.}\ \bibnamefont {Schelin}}, \bibinfo {author}
  {\bibfnamefont {J.}~\bibnamefont {Szezech}}, \bibinfo {author} {\bibfnamefont
  {D.~L.}\ \bibnamefont {Toufen}},  \emph {et~al.},\ }\bibfield  {title}
  {\enquote {\bibinfo {title} {Shearless transport barriers in magnetically
  confined plasmas},}\ }\href@noop {} {\bibfield  {journal} {\bibinfo
  {journal} {Plasma Phys. Control. Fusion}\ }\textbf {\bibinfo {volume} {54}},\
  \bibinfo {pages} {124035} (\bibinfo {year} {2012})}\BibitemShut {NoStop}%
\bibitem [{\citenamefont {Horton}(1985)}]{horton1985}%
  \BibitemOpen
  \bibfield  {author} {\bibinfo {author} {\bibfnamefont {W.}~\bibnamefont
  {Horton}},\ }\bibfield  {title} {\enquote {\bibinfo {title} {Onset of
  stochasticity and the diffusion approximation in drift waves},}\ }\href@noop
  {} {\bibfield  {journal} {\bibinfo  {journal} {Plasma Phys. Control. Fusion}\
  }\textbf {\bibinfo {volume} {27}},\ \bibinfo {pages} {937} (\bibinfo {year}
  {1985})}\BibitemShut {NoStop}%
\bibitem [{\citenamefont {Oda}\ and\ \citenamefont {Caldas}(1995)}]{oda1995}%
  \BibitemOpen
  \bibfield  {author} {\bibinfo {author} {\bibfnamefont {G.~A.}\ \bibnamefont
  {Oda}}\ and\ \bibinfo {author} {\bibfnamefont {I.~L.}\ \bibnamefont
  {Caldas}},\ }\bibfield  {title} {\enquote {\bibinfo {title} {Dimerized island
  chains in tokamaks},}\ }\href@noop {} {\bibfield  {journal} {\bibinfo
  {journal} {Chaos Solitons \& Fractals}\ }\textbf {\bibinfo {volume} {5}},\
  \bibinfo {pages} {15} (\bibinfo {year} {1995})}\BibitemShut {NoStop}%
\bibitem [{\citenamefont {{Egydio de Carvalho}}\ and\ \citenamefont {{Ozorio de
  Almeida}}(1992)}]{Carvalho1992}%
  \BibitemOpen
  \bibfield  {author} {\bibinfo {author} {\bibfnamefont {R.}~\bibnamefont
  {{Egydio de Carvalho}}}\ and\ \bibinfo {author} {\bibfnamefont {A.~M.}\
  \bibnamefont {{Ozorio de Almeida}}},\ }\bibfield  {title} {\enquote {\bibinfo
  {title} {Integrable approximation to the overlap of resonances},}\
  }\href@noop {} {\bibfield  {journal} {\bibinfo  {journal} {Physics Letters
  A}\ }\textbf {\bibinfo {volume} {162}},\ \bibinfo {pages} {457} (\bibinfo
  {year} {1992})}\BibitemShut {NoStop}%
\bibitem [{\citenamefont {Corso}\ and\ \citenamefont
  {Rizzato}(1998)}]{corso1998}%
  \BibitemOpen
  \bibfield  {author} {\bibinfo {author} {\bibfnamefont {G.}~\bibnamefont
  {Corso}}\ and\ \bibinfo {author} {\bibfnamefont {F.~B.}\ \bibnamefont
  {Rizzato}},\ }\bibfield  {title} {\enquote {\bibinfo {title} {Manifold
  reconnection in chaotic regimes},}\ }\href@noop {} {\bibfield  {journal}
  {\bibinfo  {journal} {Physical Review E}\ }\textbf {\bibinfo {volume} {58}},\
  \bibinfo {pages} {8013} (\bibinfo {year} {1998})}\BibitemShut {NoStop}%
\bibitem [{\citenamefont {{del-Castillo-Negrete}}(2000)}]{diego2000}%
  \BibitemOpen
  \bibfield  {author} {\bibinfo {author} {\bibfnamefont {D.}~\bibnamefont
  {{del-Castillo-Negrete}}},\ }\bibfield  {title} {\enquote {\bibinfo {title}
  {Chaotic transport in zonal flows in analogous geophysical and plasma
  systems},}\ }\href@noop {} {\bibfield  {journal} {\bibinfo  {journal} {Phys.
  Plasmas}\ }\textbf {\bibinfo {volume} {7}},\ \bibinfo {pages} {1702}
  (\bibinfo {year} {2000})}\BibitemShut {NoStop}%
\bibitem [{\citenamefont {Morrison}(2000)}]{morrison2000}%
  \BibitemOpen
  \bibfield  {author} {\bibinfo {author} {\bibfnamefont {P.~J.}\ \bibnamefont
  {Morrison}},\ }\bibfield  {title} {\enquote {\bibinfo {title} {Magnetic field
  lines, hamiltonian dynamics, and nontwist systems},}\ }\href@noop {}
  {\bibfield  {journal} {\bibinfo  {journal} {Phys. Plasmas}\ }\textbf
  {\bibinfo {volume} {7}},\ \bibinfo {pages} {2279} (\bibinfo {year}
  {2000})}\BibitemShut {NoStop}%
\bibitem [{\citenamefont {Horton}\ \emph {et~al.}(1998)\citenamefont {Horton},
  \citenamefont {Park}, \citenamefont {Kwon}, \citenamefont {Strozzi},
  \citenamefont {Morrison},\ and\ \citenamefont {Choi}}]{horton1998}%
  \BibitemOpen
  \bibfield  {author} {\bibinfo {author} {\bibfnamefont {W.}~\bibnamefont
  {Horton}}, \bibinfo {author} {\bibfnamefont {H.~B.}\ \bibnamefont {Park}},
  \bibinfo {author} {\bibfnamefont {J.~M.}\ \bibnamefont {Kwon}}, \bibinfo
  {author} {\bibfnamefont {D.}~\bibnamefont {Strozzi}}, \bibinfo {author}
  {\bibfnamefont {P.~J.}\ \bibnamefont {Morrison}}, \ and\ \bibinfo {author}
  {\bibfnamefont {D.~I.}\ \bibnamefont {Choi}},\ }\bibfield  {title} {\enquote
  {\bibinfo {title} {Drift wave test particle transport in reversed shear
  profile},}\ }\href@noop {} {\bibfield  {journal} {\bibinfo  {journal} {Phys.
  Plasmas}\ }\textbf {\bibinfo {volume} {5}},\ \bibinfo {pages} {3910}
  (\bibinfo {year} {1998})}\BibitemShut {NoStop}%
\bibitem [{\citenamefont {Rosalem}, \citenamefont {Roberto},\ and\
  \citenamefont {Caldas}(2014)}]{rosalem2014}%
  \BibitemOpen
  \bibfield  {author} {\bibinfo {author} {\bibfnamefont {K.~C.}\ \bibnamefont
  {Rosalem}}, \bibinfo {author} {\bibfnamefont {M.}~\bibnamefont {Roberto}}, \
  and\ \bibinfo {author} {\bibfnamefont {I.~L.}\ \bibnamefont {Caldas}},\
  }\bibfield  {title} {\enquote {\bibinfo {title} {Influence of the electric
  and magnetic shears on tokamak transport},}\ }\href@noop {} {\bibfield
  {journal} {\bibinfo  {journal} {Nucl. Fusion}\ }\textbf {\bibinfo {volume}
  {54}},\ \bibinfo {pages} {064001} (\bibinfo {year} {2014})}\BibitemShut
  {NoStop}%
\bibitem [{\citenamefont {Marcus}\ \emph {et~al.}(2019)\citenamefont {Marcus},
  \citenamefont {Roberto}, \citenamefont {Caldas}, \citenamefont {Rosalem},\
  and\ \citenamefont {Elskens}}]{marcus2019}%
  \BibitemOpen
  \bibfield  {author} {\bibinfo {author} {\bibfnamefont {F.~A.}\ \bibnamefont
  {Marcus}}, \bibinfo {author} {\bibfnamefont {M.}~\bibnamefont {Roberto}},
  \bibinfo {author} {\bibfnamefont {I.~L.}\ \bibnamefont {Caldas}}, \bibinfo
  {author} {\bibfnamefont {K.~C.}\ \bibnamefont {Rosalem}}, \ and\ \bibinfo
  {author} {\bibfnamefont {Y.}~\bibnamefont {Elskens}},\ }\bibfield  {title}
  {\enquote {\bibinfo {title} {Influence of the radial electric field on the
  shearless transport barriers in tokamaks},}\ }\href@noop {} {\bibfield
  {journal} {\bibinfo  {journal} {Phys. Plasmas}\ }\textbf {\bibinfo {volume}
  {26}},\ \bibinfo {pages} {022302} (\bibinfo {year} {2019})}\BibitemShut
  {NoStop}%
\bibitem [{\citenamefont {Osorio}\ \emph {et~al.}(2021)\citenamefont {Osorio},
  \citenamefont {Roberto}, \citenamefont {Caldas}, \citenamefont {Viana},\ and\
  \citenamefont {Elskens}}]{osorio2021}%
  \BibitemOpen
  \bibfield  {author} {\bibinfo {author} {\bibfnamefont {L.}~\bibnamefont
  {Osorio}}, \bibinfo {author} {\bibfnamefont {M.}~\bibnamefont {Roberto}},
  \bibinfo {author} {\bibfnamefont {I.~L.}\ \bibnamefont {Caldas}}, \bibinfo
  {author} {\bibfnamefont {R.~L.}\ \bibnamefont {Viana}}, \ and\ \bibinfo
  {author} {\bibfnamefont {Y.}~\bibnamefont {Elskens}},\ }\bibfield  {title}
  {\enquote {\bibinfo {title} {Onset of internal transport barriers in
  tokamaks},}\ }\href@noop {} {\bibfield  {journal} {\bibinfo  {journal} {Phys.
  Plasmas}\ }\textbf {\bibinfo {volume} {28}},\ \bibinfo {pages} {082305}
  (\bibinfo {year} {2021})}\BibitemShut {NoStop}%
\bibitem [{\citenamefont {{El Mouden}}\ \emph {et~al.}(2007)\citenamefont {{El
  Mouden}}, \citenamefont {Saifaoui}, \citenamefont {Zine},\ and\ \citenamefont
  {Eddahoy}}]{mouden2007}%
  \BibitemOpen
  \bibfield  {author} {\bibinfo {author} {\bibfnamefont {M.}~\bibnamefont {{El
  Mouden}}}, \bibinfo {author} {\bibfnamefont {D.}~\bibnamefont {Saifaoui}},
  \bibinfo {author} {\bibfnamefont {B.}~\bibnamefont {Zine}}, \ and\ \bibinfo
  {author} {\bibfnamefont {M.}~\bibnamefont {Eddahoy}},\ }\bibfield  {title}
  {\enquote {\bibinfo {title} {Transport barriers with magnetic shear in a
  tokamak},}\ }\href@noop {} {\bibfield  {journal} {\bibinfo  {journal} {J.
  Plasma Phys.}\ }\textbf {\bibinfo {volume} {73}},\ \bibinfo {pages} {439}
  (\bibinfo {year} {2007})}\BibitemShut {NoStop}%
\bibitem [{\citenamefont {Grime}\ \emph {et~al.}(2023)\citenamefont {Grime},
  \citenamefont {Roberto}, \citenamefont {Viana}, \citenamefont {Elskens},\
  and\ \citenamefont {Caldas}}]{grime2023}%
  \BibitemOpen
  \bibfield  {author} {\bibinfo {author} {\bibfnamefont {G.}~\bibnamefont
  {Grime}}, \bibinfo {author} {\bibfnamefont {M.}~\bibnamefont {Roberto}},
  \bibinfo {author} {\bibfnamefont {R.}~\bibnamefont {Viana}}, \bibinfo
  {author} {\bibfnamefont {Y.}~\bibnamefont {Elskens}}, \ and\ \bibinfo
  {author} {\bibfnamefont {I.}~\bibnamefont {Caldas}},\ }\bibfield  {title}
  {\enquote {\bibinfo {title} {Shearless bifurcations in particle transport for
  reversed-shear tokamaks},}\ }\href@noop {} {\bibfield  {journal} {\bibinfo
  {journal} {Journal of Plasma Physics}\ }\textbf {\bibinfo {volume} {89}},\
  \bibinfo {pages} {835890101} (\bibinfo {year} {2023})}\BibitemShut {NoStop}%
\bibitem [{\citenamefont {Ritz}\ \emph {et~al.}(1984)\citenamefont {Ritz},
  \citenamefont {Bengtson}, \citenamefont {Levinson},\ and\ \citenamefont
  {Powers}}]{ritz1984}%
  \BibitemOpen
  \bibfield  {author} {\bibinfo {author} {\bibfnamefont {C.~P.}\ \bibnamefont
  {Ritz}}, \bibinfo {author} {\bibfnamefont {R.~D.}\ \bibnamefont {Bengtson}},
  \bibinfo {author} {\bibfnamefont {S.}~\bibnamefont {Levinson}}, \ and\
  \bibinfo {author} {\bibfnamefont {E.}~\bibnamefont {Powers}},\ }\bibfield
  {title} {\enquote {\bibinfo {title} {Turbulent structure in the edge plasma
  of the {TEXT} tokamak},}\ }\href@noop {} {\bibfield  {journal} {\bibinfo
  {journal} {The Physics of fluids}\ }\textbf {\bibinfo {volume} {27}},\
  \bibinfo {pages} {2956} (\bibinfo {year} {1984})}\BibitemShut {NoStop}%
\bibitem [{\citenamefont {Kwon}\ \emph {et~al.}(2000)\citenamefont {Kwon},
  \citenamefont {Horton}, \citenamefont {Zhu}, \citenamefont {Morrison},
  \citenamefont {Park},\ and\ \citenamefont {Choi}}]{kwon2000}%
  \BibitemOpen
  \bibfield  {author} {\bibinfo {author} {\bibfnamefont {J.-M.}\ \bibnamefont
  {Kwon}}, \bibinfo {author} {\bibfnamefont {W.}~\bibnamefont {Horton}},
  \bibinfo {author} {\bibfnamefont {P.}~\bibnamefont {Zhu}}, \bibinfo {author}
  {\bibfnamefont {P.~J.}\ \bibnamefont {Morrison}}, \bibinfo {author}
  {\bibfnamefont {H.-B.}\ \bibnamefont {Park}}, \ and\ \bibinfo {author}
  {\bibfnamefont {D.~I.}\ \bibnamefont {Choi}},\ }\bibfield  {title} {\enquote
  {\bibinfo {title} {Global drift wave map test particle simulations},}\
  }\href@noop {} {\bibfield  {journal} {\bibinfo  {journal} {Phys. Plasmas}\
  }\textbf {\bibinfo {volume} {7}},\ \bibinfo {pages} {1169} (\bibinfo {year}
  {2000})}\BibitemShut {NoStop}%
\bibitem [{\citenamefont {Lichtenberg}\ and\ \citenamefont
  {Lieberman}(1997)}]{lichtenberg}%
  \BibitemOpen
  \bibfield  {author} {\bibinfo {author} {\bibfnamefont {A.~J.}\ \bibnamefont
  {Lichtenberg}}\ and\ \bibinfo {author} {\bibfnamefont {M.~A.}\ \bibnamefont
  {Lieberman}},\ }\href@noop {} {\enquote {\bibinfo {title} {Regular and
  chaotic motion},}\ } (\bibinfo {year} {1997})\BibitemShut {NoStop}%
\bibitem [{\citenamefont {Greene}(1979)}]{greene1979}%
  \BibitemOpen
  \bibfield  {author} {\bibinfo {author} {\bibfnamefont {J.~M.}\ \bibnamefont
  {Greene}},\ }\bibfield  {title} {\enquote {\bibinfo {title} {A method for
  determining a stochastic transition},}\ }\href@noop {} {\bibfield  {journal}
  {\bibinfo  {journal} {Journal of Mathematical Physics}\ }\textbf {\bibinfo
  {volume} {20}},\ \bibinfo {pages} {1183} (\bibinfo {year}
  {1979})}\BibitemShut {NoStop}%
\bibitem [{\citenamefont {Meiss}(1992)}]{meiss1992}%
  \BibitemOpen
  \bibfield  {author} {\bibinfo {author} {\bibfnamefont {J.~D.}\ \bibnamefont
  {Meiss}},\ }\bibfield  {title} {\enquote {\bibinfo {title} {Symplectic maps,
  variational principles, and transport},}\ }\href@noop {} {\bibfield
  {journal} {\bibinfo  {journal} {Rev. Mod. Phys.}\ }\textbf {\bibinfo {volume}
  {64}},\ \bibinfo {pages} {795} (\bibinfo {year} {1992})}\BibitemShut
  {NoStop}%
\bibitem [{\citenamefont {Viana}\ \emph {et~al.}(2021)\citenamefont {Viana},
  \citenamefont {Caldas}, \citenamefont {Szezech~Jr}, \citenamefont {Batista},
  \citenamefont {Abud}, \citenamefont {Schelin}, \citenamefont {Mugnaine},
  \citenamefont {Santos}, \citenamefont {Leal}, \citenamefont {Bartoloni} \emph
  {et~al.}}]{viana2021}%
  \BibitemOpen
  \bibfield  {author} {\bibinfo {author} {\bibfnamefont {R.}~\bibnamefont
  {Viana}}, \bibinfo {author} {\bibfnamefont {I.~L.}\ \bibnamefont {Caldas}},
  \bibinfo {author} {\bibfnamefont {J.}~\bibnamefont {Szezech~Jr}}, \bibinfo
  {author} {\bibfnamefont {A.}~\bibnamefont {Batista}}, \bibinfo {author}
  {\bibfnamefont {C.}~\bibnamefont {Abud}}, \bibinfo {author} {\bibfnamefont
  {A.}~\bibnamefont {Schelin}}, \bibinfo {author} {\bibfnamefont
  {M.}~\bibnamefont {Mugnaine}}, \bibinfo {author} {\bibfnamefont
  {M.}~\bibnamefont {Santos}}, \bibinfo {author} {\bibfnamefont
  {B.}~\bibnamefont {Leal}}, \bibinfo {author} {\bibfnamefont {B.}~\bibnamefont
  {Bartoloni}},  \emph {et~al.},\ }\bibfield  {title} {\enquote {\bibinfo
  {title} {Transport barriers in symplectic maps},}\ }\href@noop {} {\bibfield
  {journal} {\bibinfo  {journal} {Brazilian Journal of Physics}\ }\textbf
  {\bibinfo {volume} {51}},\ \bibinfo {pages} {899} (\bibinfo {year}
  {2021})}\BibitemShut {NoStop}%
\bibitem [{\citenamefont {Szezech~Jr}\ \emph {et~al.}(2009)\citenamefont
  {Szezech~Jr}, \citenamefont {Caldas}, \citenamefont {Lopes}, \citenamefont
  {Viana},\ and\ \citenamefont {Morrison}}]{szezech2009}%
  \BibitemOpen
  \bibfield  {author} {\bibinfo {author} {\bibfnamefont {J.~D.}\ \bibnamefont
  {Szezech~Jr}}, \bibinfo {author} {\bibfnamefont {I.~L.}\ \bibnamefont
  {Caldas}}, \bibinfo {author} {\bibfnamefont {S.~R.}\ \bibnamefont {Lopes}},
  \bibinfo {author} {\bibfnamefont {R.~L.}\ \bibnamefont {Viana}}, \ and\
  \bibinfo {author} {\bibfnamefont {P.~J.}\ \bibnamefont {Morrison}},\
  }\bibfield  {title} {\enquote {\bibinfo {title} {Transport properties in
  nontwist area-preserving maps},}\ }\href@noop {} {\bibfield  {journal}
  {\bibinfo  {journal} {Chaos}\ }\textbf {\bibinfo {volume} {19}},\ \bibinfo
  {pages} {043108} (\bibinfo {year} {2009})}\BibitemShut {NoStop}%
\bibitem [{\citenamefont {Prince}\ and\ \citenamefont
  {Dormand}(1981)}]{prince1981}%
  \BibitemOpen
  \bibfield  {author} {\bibinfo {author} {\bibfnamefont {P.~J.}\ \bibnamefont
  {Prince}}\ and\ \bibinfo {author} {\bibfnamefont {J.~R.}\ \bibnamefont
  {Dormand}},\ }\bibfield  {title} {\enquote {\bibinfo {title} {High order
  embedded {R}unge-{K}utta formulae},}\ }\href@noop {} {\bibfield  {journal}
  {\bibinfo  {journal} {Journal of computational and applied mathematics}\
  }\textbf {\bibinfo {volume} {7}},\ \bibinfo {pages} {67} (\bibinfo {year}
  {1981})}\BibitemShut {NoStop}%
\bibitem [{\citenamefont {Nascimento}\ \emph {et~al.}(2005)\citenamefont
  {Nascimento}, \citenamefont {Kuznetsov}, \citenamefont {Severo},
  \citenamefont {Fonseca}, \citenamefont {Elfimov}, \citenamefont {Bellintani},
  \citenamefont {Machida}, \citenamefont {Heller}, \citenamefont {Galv\~{a}o},\
  and\ \citenamefont {Sanada}}]{nascimento2005}%
  \BibitemOpen
  \bibfield  {author} {\bibinfo {author} {\bibfnamefont {I.~C.}\ \bibnamefont
  {Nascimento}}, \bibinfo {author} {\bibfnamefont {Y.~K.}\ \bibnamefont
  {Kuznetsov}}, \bibinfo {author} {\bibfnamefont {J.~H.~F.}\ \bibnamefont
  {Severo}}, \bibinfo {author} {\bibfnamefont {A.~M.~M.}\ \bibnamefont
  {Fonseca}}, \bibinfo {author} {\bibfnamefont {A.}~\bibnamefont {Elfimov}},
  \bibinfo {author} {\bibfnamefont {V.}~\bibnamefont {Bellintani}}, \bibinfo
  {author} {\bibfnamefont {M.}~\bibnamefont {Machida}}, \bibinfo {author}
  {\bibfnamefont {M.~V. A.~P.}\ \bibnamefont {Heller}}, \bibinfo {author}
  {\bibfnamefont {R.~M.~O.}\ \bibnamefont {Galv\~{a}o}}, \ and\ \bibinfo
  {author} {\bibfnamefont {E.~K.}\ \bibnamefont {Sanada}},\ }\bibfield  {title}
  {\enquote {\bibinfo {title} {Plasma confinement using biased electrode in the
  {TCABR} tokamak},}\ }\href@noop {} {\bibfield  {journal} {\bibinfo  {journal}
  {Nucl. Fusion}\ }\textbf {\bibinfo {volume} {45}},\ \bibinfo {pages} {796}
  (\bibinfo {year} {2005})}\BibitemShut {NoStop}%
\bibitem [{\citenamefont {Marcus}\ \emph {et~al.}(2008)\citenamefont {Marcus},
  \citenamefont {Caldas}, \citenamefont {{Guimar{\~a}es-Filho}}, \citenamefont
  {Morrison}, \citenamefont {Horton}, \citenamefont {Kuznetsov},\ and\
  \citenamefont {Nascimento}}]{marcus2008}%
  \BibitemOpen
  \bibfield  {author} {\bibinfo {author} {\bibfnamefont {F.~A.}\ \bibnamefont
  {Marcus}}, \bibinfo {author} {\bibfnamefont {I.~L.}\ \bibnamefont {Caldas}},
  \bibinfo {author} {\bibfnamefont {Z.~O.}\ \bibnamefont
  {{Guimar{\~a}es-Filho}}}, \bibinfo {author} {\bibfnamefont {P.~J.}\
  \bibnamefont {Morrison}}, \bibinfo {author} {\bibfnamefont {W.}~\bibnamefont
  {Horton}}, \bibinfo {author} {\bibfnamefont {Y.~K.}\ \bibnamefont
  {Kuznetsov}}, \ and\ \bibinfo {author} {\bibfnamefont {I.~C.}\ \bibnamefont
  {Nascimento}},\ }\bibfield  {title} {\enquote {\bibinfo {title} {Reduction of
  chaotic particle transport driven by drift waves in sheared flows},}\
  }\href@noop {} {\bibfield  {journal} {\bibinfo  {journal} {Physics of
  Plasmas}\ }\textbf {\bibinfo {volume} {15}},\ \bibinfo {pages} {112304}
  (\bibinfo {year} {2008})}\BibitemShut {NoStop}%
\bibitem [{\citenamefont {Severo}\ \emph {et~al.}(2021)\citenamefont {Severo},
  \citenamefont {Canal}, \citenamefont {Ronchi}, \citenamefont {Andrade},
  \citenamefont {Fernandes}, \citenamefont {Ikeda}, \citenamefont {Collares},
  \citenamefont {Galv{\~a}o}, \citenamefont {Nascimento},\ and\ \citenamefont
  {Tendler}}]{severo2021}%
  \BibitemOpen
  \bibfield  {author} {\bibinfo {author} {\bibfnamefont {J.~H.~F.}\
  \bibnamefont {Severo}}, \bibinfo {author} {\bibfnamefont {G.~P.}\
  \bibnamefont {Canal}}, \bibinfo {author} {\bibfnamefont {G.}~\bibnamefont
  {Ronchi}}, \bibinfo {author} {\bibfnamefont {N.~B.}\ \bibnamefont {Andrade}},
  \bibinfo {author} {\bibfnamefont {T.}~\bibnamefont {Fernandes}}, \bibinfo
  {author} {\bibfnamefont {M.~Y.}\ \bibnamefont {Ikeda}}, \bibinfo {author}
  {\bibfnamefont {M.~P.}\ \bibnamefont {Collares}}, \bibinfo {author}
  {\bibfnamefont {R.~M.~O.}\ \bibnamefont {Galv{\~a}o}}, \bibinfo {author}
  {\bibfnamefont {I.~C.}\ \bibnamefont {Nascimento}}, \ and\ \bibinfo {author}
  {\bibfnamefont {M.}~\bibnamefont {Tendler}},\ }\bibfield  {title} {\enquote
  {\bibinfo {title} {Overview of plasma rotation studies on the {TCABR}
  tokamak},}\ }\href@noop {} {\bibfield  {journal} {\bibinfo  {journal} {Plasma
  Physics and Controlled Fusion}\ }\textbf {\bibinfo {volume} {63}},\ \bibinfo
  {pages} {075001} (\bibinfo {year} {2021})}\BibitemShut {NoStop}%
\bibitem [{\citenamefont {Das}\ and\ \citenamefont {Yorke}(2018)}]{das2018}%
  \BibitemOpen
  \bibfield  {author} {\bibinfo {author} {\bibfnamefont {S.}~\bibnamefont
  {Das}}\ and\ \bibinfo {author} {\bibfnamefont {J.~A.}\ \bibnamefont
  {Yorke}},\ }\bibfield  {title} {\enquote {\bibinfo {title} {Super convergence
  of ergodic averages for quasiperiodic orbits},}\ }\href@noop {} {\bibfield
  {journal} {\bibinfo  {journal} {Nonlinearity}\ }\textbf {\bibinfo {volume}
  {31}},\ \bibinfo {pages} {491} (\bibinfo {year} {2018})}\BibitemShut
  {NoStop}%
\bibitem [{\citenamefont {Das}\ \emph {et~al.}(2017)\citenamefont {Das},
  \citenamefont {Saiki}, \citenamefont {Sander},\ and\ \citenamefont
  {Yorke}}]{das2017}%
  \BibitemOpen
  \bibfield  {author} {\bibinfo {author} {\bibfnamefont {S.}~\bibnamefont
  {Das}}, \bibinfo {author} {\bibfnamefont {Y.}~\bibnamefont {Saiki}}, \bibinfo
  {author} {\bibfnamefont {E.}~\bibnamefont {Sander}}, \ and\ \bibinfo {author}
  {\bibfnamefont {J.~A.}\ \bibnamefont {Yorke}},\ }\bibfield  {title} {\enquote
  {\bibinfo {title} {Quantitative quasiperiodicity},}\ }\href@noop {}
  {\bibfield  {journal} {\bibinfo  {journal} {Nonlinearity}\ }\textbf {\bibinfo
  {volume} {30}},\ \bibinfo {pages} {4111} (\bibinfo {year}
  {2017})}\BibitemShut {NoStop}%
\bibitem [{\citenamefont {Sander}\ and\ \citenamefont
  {Meiss}(2020)}]{sander2020}%
  \BibitemOpen
  \bibfield  {author} {\bibinfo {author} {\bibfnamefont {E.}~\bibnamefont
  {Sander}}\ and\ \bibinfo {author} {\bibfnamefont {J.}~\bibnamefont {Meiss}},\
  }\bibfield  {title} {\enquote {\bibinfo {title} {Birkhoff averages and
  rotational invariant circles for area-preserving maps},}\ }\href@noop {}
  {\bibfield  {journal} {\bibinfo  {journal} {Physica D: Nonlinear Phenomena}\
  }\textbf {\bibinfo {volume} {411}},\ \bibinfo {pages} {132569} (\bibinfo
  {year} {2020})}\BibitemShut {NoStop}%
\bibitem [{\citenamefont {Wesson}\ and\ \citenamefont
  {Campbell}(2011)}]{wesson2011}%
  \BibitemOpen
  \bibfield  {author} {\bibinfo {author} {\bibfnamefont {J.}~\bibnamefont
  {Wesson}}\ and\ \bibinfo {author} {\bibfnamefont {D.~J.}\ \bibnamefont
  {Campbell}},\ }\href@noop {} {\enquote {\bibinfo {title} {Tokamaks},}\ }
  (\bibinfo {year} {2011})\BibitemShut {NoStop}%
\end{thebibliography}%

\end{document}